\documentclass[useAMS,usenatbib]{mnras}

\usepackage{graphicx}
\usepackage{longtable}
\usepackage{amssymb}
\usepackage{booktabs}
\usepackage{tabularx, blindtext}

\title[MIFAL: automated multiple image finder]{ MIFAL: fully automated Multiple-Image Finder ALgorithm for strong-lens modelling -- proof of concept}
\author[Carrasco, Zitrin \& Seidel]{Mauricio Carrasco$^{1}$, Adi Zitrin$^{2}$\thanks{E-mail: adizitrin@gmail.com} \& Gregor Seidel$^{3}$ \\\\
$^{1}$Universit\"at Heidelberg, Zentrum f\"ur Astronomie, Institut f\"ur Theoretische Astrophysik, Philosophenweg 12, 69120 Heidelberg, Germany\\
$^{2}$Department of Physics, Ben-Gurion University of the Negev, Be'er-Sheva 84105, Israel\\
$^{3}$Max-Planck-Institute for Astronomy, K\"onigstuhl 17, D-69117 Heidelberg, Germany}

\begin{document}


\pagerange{\pageref{firstpage}--\pageref{lastpage}} \pubyear{2013}

\maketitle

\label{firstpage}

\begin{abstract}
We outline a simple procedure designed for \emph{automatically} finding sets of multiple images in strong lensing (SL) clusters. 
We show that by combining (a) an arc-finding (or source extracting) program, (b) photometric redshift measurements, and (c) a preliminary light-traces-mass lens model, multiple-image systems can be identified in a fully automated (``blind'') manner. The presented procedure yields an assessment of the likelihood of each arc to belong to one of the multiple-image systems, as well as the preferred redshift for the different systems. These could be then used to automatically constrain and refine the 
initial lens model for an accurate mass distribution. We apply this procedure to \emph{Cluster Lensing And Supernova with Hubble} observations of three galaxy clusters, MACS J0329.6-0211, MACS J1720.2+3536, and MACS J1931.8-2635, comparing the results to published SL analyses where multiple images were verified by eye on a particular basis. In the first cluster all originally identified systems are recovered by the automated procedure, and in the second and third clusters about half are recovered. Other known systems are not picked up, in part due to a crude choice of parameters, ambiguous photometric redshifts, or inaccuracy of the initial lens model. On top of real systems recovered, some false images are also mistakenly identified by the procedure, depending on the thresholds used. While further improvements to the procedure and a more thorough scrutinisation of its performance are warranted, the work constitutes another important step toward fully automatising SL analyses for studying mass distributions of large cluster samples.
\end{abstract}


\section{Introduction}\label{intro}
The matter in the Universe is thought to consist of mostly an unseen component, known as Dark Matter (DM; e.g. \citealt{Komatsu2011WMAP7,Planck2018Params}). Although DM has still not been detected directly, evidence for it existence is manifested in many astrophysical phenomena and measurements on various scales \citep[][for a review]{Einasto2009DMreview}.

One of the most direct ways to map the unseen DM distribution in massive objects such as galaxy clusters is through the analysis of 
gravitational lensing signatures. Due to the high projected mass density of clusters, background sources are lensed, typically forming magnified and distorted images. 
Close to the cluster centre, in the strong-lensing (SL) regime where the mass density is higher than the critical value for lensing (see \citealt{Bartelmann2010reviewB,Kneib2011review}), there are often seen \emph{multiple} images of the same background source.  As lensing is coupled to the deflecting mass distribution, the location and redshift of the multiple images supply constraints to map the DM distribution. 
Matching up multiple-image families, often acknowledged as the most complicated and time-consuming step in a SL analysis 
\cite[see, e.g.][]{Jullo2007Lenstool,Bradac2008rxj1347}, is thus a crucial step for the modelling of the inner cluster (unseen) mass distribution.

Methods for analysing strong-lenses have been evolving over the past decades in response to higher quality data, mainly from space, and increasing computational power
(e.g., \citealt{Kneib1993Lenstool,Keeton2001models,Broadhurst2005a,Halkola2006,Liesenborgs2006,Jullo2007Lenstool,Coe2008LensPerfect},
see also \citealt{Kneib2011review}). Moreover, extensive lensing surveys recently undertaken with the \emph{Hubble Space Telescope (HST)} such as the Cluster Lensing and Supernova Survey with Hubble (CLASH; \citealt{PostmanCLASHoverview}), the Reionization Lensing Cluster Survey (RELICS; \citealt{Coe2019arXivRELICS}) and the ultra-deep Hubble Frontier Fields \citep{Lotz2016HFF}, have allowed for high-end modelling using large numbers of multiple images \citep[e.g.][]{Jauzac2015A2744,Caminha2017M0416}, and for the comparison of different lens modelling techniques \citep{Treu2016Refsdal}, including with simulations \citep{Meneghetti2016SIMSCOMP}. Methods for SL analyses generally divide into two categories, namely \emph{parametric} and so-called \emph{non-parametric} or \emph{free-form} methods. Free-form
techniques (e.g., \citealt{Diego2005Nonparam,Liesenborgs2007,Coe2010,WilliamsSaha2011Offset}) usually make no assumptions regarding the underlying distribution of matter, 
and directly constrain the mass distribution from the location and redshift of multiple-images. Although this procedure is appealing due to its 
model-independency, the number of multiple-images readily identified in lensing clusters is often too small to obtain a 
high-resolution result \citep{PonenteDiego2011}. For this reason, most free-form techniques often do not possess the capability to predict many multiple images in terms of location, exact shape, internal details and orientation (although few can also be successful 
in doing so, for example see \citealt{Liesenborgs2008CL0024L} or when combining the free-form method with the smaller-scale parametric galaxy contributions, e.g., \citealt{Diego2014M0717,Diego2016refsdal}). Parametric techniques \citep[e.g.][]{Jullo2007Lenstool,Johnson2014HFFmodels,Oguri2012SL,Grillo2015_0416},
 on the other hand, make certain well-motivated assumptions regarding the underlying mass distribution and use physical parametrisations to represent it, and when
 well constructed, could be used to successfully find multiple-images and accurately predict the location and shape of other counter images \citep[e.g.][]{Treu2016Refsdal,Meneghetti2016SIMSCOMP}.

In fact, this prediction power is the key point for the work presented here. Based on previous studies in which we have analysed a few dozen galaxy clusters
\citep[e.g.][]{Zitrin2009b,Zitrin2015CLASH25}, we noticed that by constructing a simple 
\emph{preliminary} yet well-guessed lens model, multiple-images can be readily found using the model, by sending manually each arc-like image to the source plane and back through the lens to the image plane, and then looking for similar looking (and similarly redshifted) objects where the model predicts them.
 Thus, our goal here is to simply automatise this procedure, so that clusters could be analysed ``blindly'', in an automated manner.

Despite the various methods available for SL analyses or mass modelling, only several dozen clusters have been analysed to date with regards
 to their SL features (albeit with rapidly increasing numbers), with the vast majority only in the last decade or so
\citep[e.g.][]{Smith2005,Richard2010locuss20,Zitrin2015CLASH25,Kawamata2016modelsHFF,Cerny2018,Acebron2018}. One of the main reasons, apart from the finite amount of 
high-resolution data particularly from space, is that nearly all methods are often time-consuming and the analysis has to be performed on a particular
 basis with significant intervention and fine-tuning by the user. Ongoing large sky-surveys and the amount of data taken by high-resolution space 
missions such as the \emph{HST}, the upcoming launch of the \emph{James Webb Space Telescope (JWST)}, and especially the \emph{Euclid} and  \emph{WFIRST} missions, call for a more (statistically) efficient and automated method which could be operated on a broad database. 

Broad studies of galaxy clusters are of high importance, as they will be able to probe and put independent constraints on the acceptable $\Lambda$CDM paradigm \citep[e.g.][]{Wambsganss1995Testcdm,Dalal+2004arcs,OguriBlandford2009}, which is often confronted by various studies \citep[e.g.][see also \citealt{Kroupa2012problemDM}; although most show a mild discrepancy with debatable 
consequences]{Hennawi2007,Broadhurst2008,BroadhurstBarkana2008,SadehRephaeli2008,PuchweinHilbert2009,
Sereno2010,Gralla2011,Meneghetti2010b,Umetsu2011}, and on related and important observables such as the geometry 
of the Universe \citep[e.g.][]{Jullo2010}, the mass-concentration relation of clusters \citep[e.g.][see also \citealt{vonderLinden2014WTG, Hoekstra2015CCCP, Okabe2016cM}]{Oguri201238clusters,Merten2014CLASHcM,Umetsu2014CLASH_WL}, the 
density profiles in the inner core \citep{Navarro1996,Newman2013}, the high-end of the mass function and its evolvement in redshift, or the
 (in)homogeneity of the Universe. This is especially relevant given the recent tension between the abundance of massive clusters found by the \emph{Planck} satellite, and the \emph{Planck} CMB cosmology \citep{Planck2015catalog}. As signatures of lensing are coupled, via the mass distribution, to the gravitational potential, it is also one of the very few phenomena that \emph{might}, conceivably, hold the key for independently testing the nature of DM, or its alternatives such as MOND or TeVes/MOG (e.g., \citealt{Milgrom1983MOND,Bekenstein2012ReviewTeVes}, see also \citealt{Amendola2008,Tian2012TeVesLensing}), Early 
Dark Energy propositions \citep[see][]{Fedeli2007EDE1} and other innovative ideas such as the \emph{large local void} in a
 Lema\^itre-Tolman-Bondi (LTB) universe \citep[see][]{Redlich2014LTB}, or non-Gaussianity of early 
perturbations \citep{ChongchitnanSilk2012}, and free-streaming neutrinos \citep{Emami2017NeutrinosClusters}.

In addition, since galaxy clusters magnify background sources due to the lensing effect, lens models and their magnification maps are central in the discovery 
and higher-resolution studies of early, high-$z$ galaxies \citep[e.g.][]{Bradley2011,Bouwens2009behindclusters,Zheng2009,
Zackrisson2012,Bradac2012highz,Ishigaki2014,Hashimoto2018}. In particular, recent deep observations along with successful mass modelling, enable also the 
detection of \emph{multiply-imaged}, higher-redshift galaxies \citep[e.g.][]{Franx1997,Kneib2004z7,Richard2011,Coe2012highz}.

Here, we combine three successful (and independent) tools, with the goal of ``paving the road'' towards a completely automated and innovative 
(albeit simple) multiple-image finder, or equivalently, SL analyser. We begin with a sophisticated arc-finder \citep{Seidel2007Arcfinder} 
to detect stretched and arc-shaped objects across the cluster central field (that are likely lensed and, potentially, multiply imaged). The photometry of these arcs could then either be input into a photometric-redshift code (such as the Bayesian Photometric Redshift, BPZ; \citealt{Benitez2000, Benitez2004, Coe2006}) in order to assess the corresponding source redshift, or in our case, cross-matched with objects in the public CLASH catalogues, which include multi-band photometry and photometric redshift estimates for each object. We construct an  initial mass model using a revised version of the (semi-)parametric light-traces-mass (LTM) SL modelling method of \citet{Zitrin2009b}, which has a minimum of free parameters and is initially well constrained even without using multiple-images as input, and is therefore optimal for finding and confirming new multiple-images
 \citep[e.g.][]{Zitrin2015CLASH25}. We make use of the initial mass model to \emph{automatically} lens each arc candidate back to the source
 plane via the lens equation, with the measured photometric redshift, and relens it through the lens to predict the appearance of other counter images
 of the same background source. In each of the predicted locations in the image-plane, the arc catalogue is searched for close-by objects with similar 
photometric redshifts, surface brightness, and colours. The new multiple-image families thus found can then, possibly iteratively, be used in order to refine the mass model. Although it constitutes another important step towards automated lensing analyses potentially applicable to high-end data or 
large sky surveys \citep[e.g.][]{Zitrin2011d,Wong2012OptLenses,Stapelberg2019}, the outlined procedure is clearly simple in essence. Also, each of the three independent ingredients of the algorithm could be in principle easily replaced with other alternatives, or future improvements.

We apply here this method to the X-ray luminous \citep{Ebeling2010FinalMACS} clusters MACS J0329.6-0211,  MACS J1720.2+3536, and MACS J1931.8-2635 (hereafter M0329, M1720, and M1931, respectively), in observations taken with the \emph{HST} as part of the CLASH treasury program \citep{PostmanCLASHoverview}. The clusters were imaged to a
depth of $\sim 27$ AB in 16 filters with the Wide Field Camera 3 (WFC3) UVIS and IR cameras, and the Advanced Camera for Surveys
(ACS) WFC. The images were reduced and mosaiced ($0.065 \arcsec/pix$) using standard techniques implemented in the MosaicDrizzle pipeline
\citep{Koekemoer2002,Koekemoer2011}. CLASH also delivered HST photometry for all identified objects performed automatically by SExtractor \citep{BertinArnouts1996Sextractor}, and accompanying photometric redshifts estimated for all the galaxies in the field using the full 16-band UVIS/ACS/WFC3-IR photometry via the BPZ \citep{Benitez2000,Benitez2004,Coe2006} program (\citealt{PostmanCLASHoverview,Molino2017CLASHcats} for additional details).  

The paper is organised as follows: In \S \ref{algo} we detail the multiple-image finder and SL analysis algorithms. In \S \ref{SLanalysis} we report the results of applying
this procedure to M0329, M1720, and M1931, and compare the results with previous analyses. The results are discussed in \S \ref{Discussion} and the work is summarised in \S \ref{summary}. The code to construct an initial blind model is given in Appendix \ref{appendix:1}, and the relevant parameters that were used are given in Appendix \ref{appendix:2}. Throughout this paper we adopt a concordance $\Lambda$CDM cosmology with ($\Omega_{\rm m0}=0.3$, $\Omega_{\Lambda 0}=0.7$, $H_{0}=100$ $h$ km s$^{-1}$Mpc$^{-1}$, with $h=0.7$).

\begin{figure*}
 \begin{center}
   \includegraphics[width=140mm]{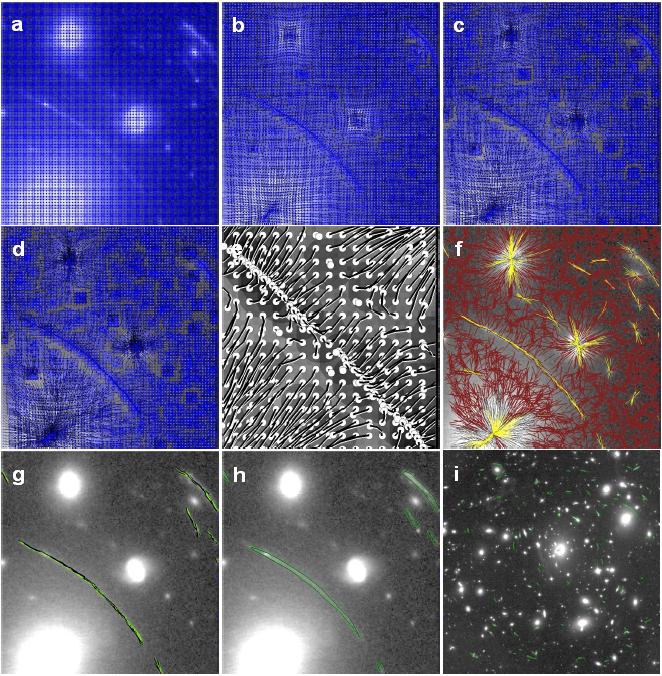}
 \end{center}
\caption{Example of some key steps in the arcfinder algorithm. The upper-left panel (a) shows the initial division into cells in an equally spaced grid. The 
following sub figures (``b-e'') show the iterative displacement of the cells towards their centres of brightness. ``e'' shows a zoom-in of the complete
 process where the black lines trace the path of each cell, shown by a white circle. Then, highly correlated cells are searched for, marked 
in yellow in sub figure ``f''. These ensembles of cells are now combined into objects using a friends-of-friends type algorithm, and the resulting objects
 are shown in ``g'', where their calculated brightness contours are shown in ``h''. Sub figure ``i'' shows the central field of M0329 and the arclets found
 therein in green contours (shown also in Fig. \ref{C9-fig:greg2_zoom} for a better view).}
\label{C9-fig:greg1}
\end{figure*}

\begin{figure*}
 \begin{center}
   \includegraphics[width=120mm]{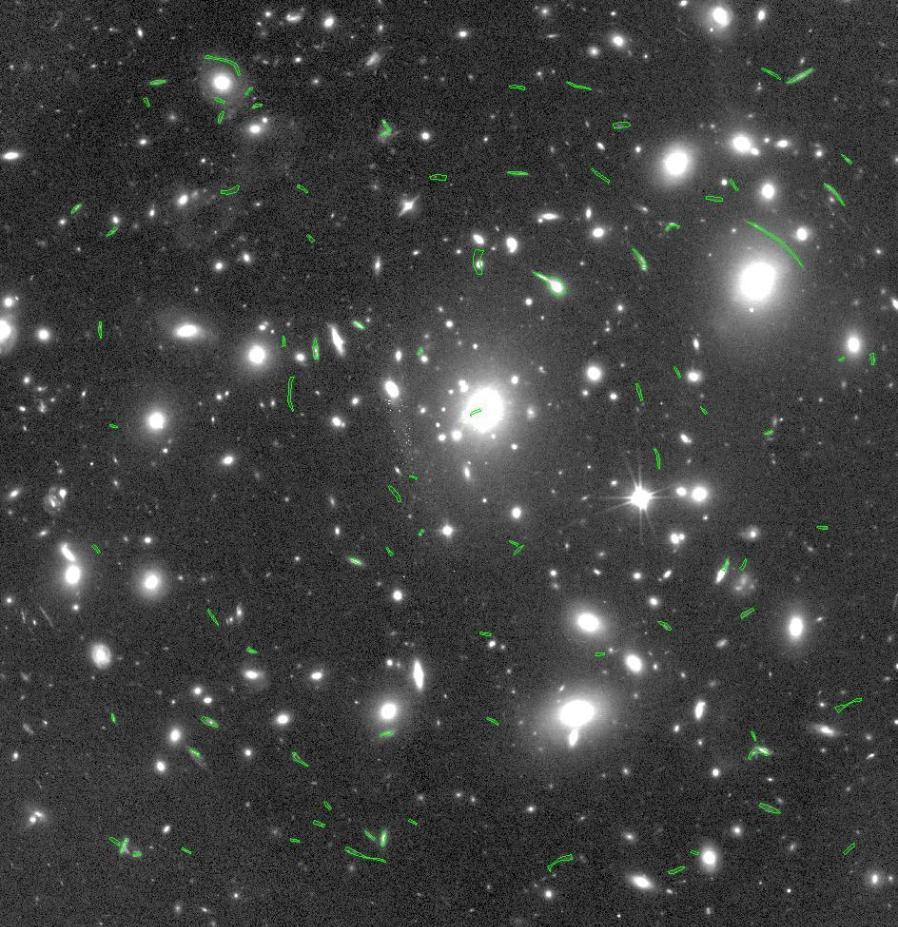}
 \end{center}
\caption{Same as Fig. 1i, but enlarged for better view.}
\label{C9-fig:greg2_zoom}
\end{figure*}

\section{The Multiple-Image Finder ALgorithm}\label{algo}
We present here the \textbf{M}ultiple-\textbf{I}mage \textbf{F}inder \textbf{AL}gorithm (MIFAL)\footnote{Based on work presented in M. Carrasco's PhD Dissertation}. The general idea is as follows: given (i) a list of arclet candidates near the cluster core, (ii) their photometry and photometric redshifts, and (iii) a preliminary lens model, run over all arclets, send each one to the source plane and back to obtain predicted locations for counter images, and search around these positions for objects with similar photometric redshift, Spectral Energy Distribution (SED) or colours, and surface brightness. Rank the potential candidates for each lensed arc by similarity, and select the best combinations.  We now detail the different ingredients of the algorithm.

\subsection{The arc candidate catalogue}\label{arc_cat}

The first step we apply here for automatically searching for multiple images is to primarily identify as many lensed galaxy candidates as possible
in the cluster lens field. Since lensed features are typically sheared, the first step is to build an arc catalogue, which constitutes a list of likely lensed candidates. This is the first ingredient needed to run MIFAL.

The arc catalogue is constructed here using the \emph{Arcfinder} algorithm \citep{Seidel2007Arcfinder}, which employs the first and second brightness 
moments for the detection  of elongated objects across the frame. As detailed in \citet{Seidel2007Arcfinder}, the algorithm is sufficiently robust to detect such features even if their
 surface brightness is near the pixel noise of the image, yet the amount of spurious detections remains (relatively) low.

The algorithm subdivides the image into a grid of overlapping, circular cells (see Fig. \ref{C9-fig:greg1}a) which are iteratively shifted towards their local centre of brightness
 in their immediate neighborhood (see Fig. \ref{C9-fig:greg1}b to \ref{C9-fig:greg1}e). The centre of brightness is defined as the first moment $\int_A\vec{x}d^2x = \bar{\vec{x}}$, where $A$ is 
the cell area and $q(I)$ is a weighting function that depends on the image intensity $I(\vec{x})$. The code then computes the cell ellipticity 
$(Q_{11}-Q_{22}+2 i Q_{12}) / (Q_{11}+Q_{22})$ using their second brightness moments
\begin{equation}\label{eq:Qij}
Q_{ij}=\frac{\int_{A}(x_i-\bar{x}_i)(x_j-\bar{x}_j)q(I(\vec{x}))\,d^2x}
              {\int_{A}q(I(\vec{x}))\,d^2x}.
\end{equation}
The ellipticity of each cell provides a natural measure of its orientation in the image and allows us to compute their angular separation. Cells oriented in
 the same direction, and spatially aligned, are then combined into initial objects using a simple coherence measure (see Fig. \ref{C9-fig:greg1}f); 
 essentially the product of the 
cosine of the angle between cell orientations and a (scaled) perpendicular cell separation, with a friends-of-friends type algorithm. At this point, the 
objects are nothing more than sets of correlated cells (see \ref{C9-fig:greg1}g) where one cannot directly infer much morphological information, and there are still many 
spurious detections, e.g. for spatially connected point sources. A large number of spurious detections is removed with a filter interpreting the brightness
 distribution in each cell to remove those which are unlikely to belong to an elongated object. For objects still containing a sufficient number of valid 
cells, several consecutive steps compute isophote contours which allow the computation of basic properties like length, length-to-width and signal-to-noise.
 These are then used to choose the most likely arc-shaped objects, and  further reduce the number of spurious detections. Some further filtering and noise 
cleaning procedures take place, for example to remove elongated objects which are clearly not gravitational arcs (such as diffraction spikes, spiral galaxy 
arms, edge-on galaxies, etc.), resulting in a final catalogue of \emph{arc} candidates (see \ref{C9-fig:greg1}i).

To obtain an arc catalogue as complete as possible, we run the Arcfinder on a detection image constructed from all optical ACS and WFC3-IR data. It should also be noted that the arc detection part takes only several seconds on a standard CPU. 

Two main caveats follow the use of an automated arcfinder compared to examining each arc candidate by eye. First, some spurious detections of elongated, un-lensed galaxies may appear in the catalogue. These, however,  will be mostly filtered out at a later stage based on their photometric redshifts and on the mass model. Second, images with a feeble elongation
or distortion may not be detected even if they are multiply lensed (see example the multiply-imaged spiral galaxy in MACS J1149.5+2223, 
\citealt{ZitrinBroadhurst2009}). Our goal throughout, though, is not to detect \emph{all} previously known multiple-images, but rather, 
 enough of them so that a proper mass model could be automatically constructed.

\subsection{Photometry and photometric redshifts} \label{z_phot}

The second step in the procedure is to obtain, for each arc candidate, fluxes in each band and a photometric redshift estimate. In principle, this can be done by measuring the fluxes using the Arcfinder contours directly, although here we take on an alternative route, matching each arc in the arc-candidate catalogue to an object from the public CLASH catalogues by \citet{Molino2017CLASHcats}, which include aperture-matched PSF-corrected multi-band photometry, and enhanced Bayesian Photometric Redshifts (BPZ2.0; \citealt{Benitez2000,Benitez2004,Coe2006,Molino2017CLASHcats}), based on the multi-drizzled (0.065 $\arcsec/pix$) science images produced by CLASH \citep{Koekemoer2011}. These redshifts will later be used to compute the (relative) lensing distances to re-lens the candidate arcs through the mass model and potentially, further constrain the model in the cases where multiple image systems were found. After matching arcs to their photometry and photometric redshifts, the arc catalogue is then cleaned from objects whose photometric redshift is lower than $z_l$  + 0.3, where $z_l$ is the redshift of the cluster, and from objects lying outside the field-of-view (FOV) of the initial lens model.

\subsection{The preliminary lens model} \label{pre_mass_model} 
The third and last ingredient needed to run MIFAL is a lens model. The preliminary mass model we use here is based on the semi-parametric method of \citep[][see also \citealt{Broadhurst2005a}]{Zitrin2009b,Zitrin2015CLASH25},  which is based on the assumption that light traces mass (hereafter LTM). The mass model can be quickly constructed by  adopting a simple representation of the cluster member galaxies and the underlying DM component, together with a prior guess of the ratio between the two components, and a certain (effective) mass-to-light ratio ($M/L$) to account for the overall normalisation. Details of this mass modelling method and it prediction power can be found in, e.g.,  \citet{Zitrin2009b,Zitrin2011d,Zitrin2015CLASH25}. Explicit details regarding the code and its implementation are given in Appendix \ref{appendix:1}, and here we give only a brief overview.

The mass model consists of three components. The first component are the cluster galaxies, whose surface mass density is modelled as a  power-law of slope $q$, scaled by their luminosity. The superposition of the mass density contributions from all red-sequence galaxies (brighter than a certain  threshold), represents the lumpy, galaxy component. All galaxies are modelled as circular, aside from one or two Brightest Cluster Galaxies, for which ellipticity, and potentially a core, can be assigned. This mass distribution is then smoothed using a 2D Gaussian kernel in Fourier space, to provide a model for the DM distribution in the cluster halo, which is  the second component of the model. The deflection fields from these two components are then normalised onto a similar arbitrary scale, and added together with a relative weight ($K_{gal}$). The resulting deflection field is calculated over the input image grid (or a lower resolution version of it) and scaled as desired (with parameter $K$). Finally, the third component is an external shear which can be added to the deflection field for further flexibility.  In all, there are six free parameters: the power law of the galaxies' density profile, the degree of smoothing (Gaussian width), the ratio between the galaxy and DM contributions, the overall normalisation (i.e., the mass scaling of both the smooth and galaxy components), and the strength and direction of the external shear. This simple procedure seems to produce relatively well-guessed initial mass models that can be used to identify multiple images, and in turn, be iteratively refined. This makes the procedure ideal for our purpose. 

Our goal here is to first construct a reliable initial mass model to allow for the automated identification of multiple images.  For that purpose, as we now detail, we fix all free parameters to either averaged values, or to values indicated by the input galaxy catalogue.  As we have mentioned in previous work \citep[e.g.][]{Zitrin2009b}, the choice of the power-law ($q$) and smoothing Gaussian width ($S$) only mildly affects the reproduction of multiple images. These parameters mainly affect the mass \emph{profile}, so that the assumption is that nearly all systems can, in principle, be reproduced with any reasonable choice of $q$ and $S$. Therefore we set these two parameters to typical values deduced from our previous analyses (see also \citealt{Zitrin2011d}), adopting here a power-law $q$=1.3,  and a smoothing Gaussian width of $S$=150 pixels (0.065 \arcsec/pix). Similarly, for the relative galaxy to DM ratio, we use a typical value of $K_{gal}=0.07$. The value of the overall normalisation, $K$, basically determines to which source redshift the model is scaled, and is unknown. Instead we can adopt a typical value, and iterate on different cosmological distances ($D_{ls}/D_s$ normalisations) searching for multiple images given each normalisation, and then use the multiple images found to normalise the model. Moreover, the overall normalisation $K$ correlates empirically with the size of the lens (Taiber et al., \emph{in preparation}). We use a typical value of $K=1.5$ , which -- based on previous analyses -- corresponds roughly to an Einstein radius of $\sim 20 \arcsec$. A moderate external shear ($\gamma_{ex}=0.1$) is used along the BCG direction, in cases where the BCG is highly elliptical; otherwise no external shear is used in the preliminary model. If external shear as such is used, which adds to the deflection field, then a smaller $K=1.2$ is adopted (after rescaling of the galaxy and DM deflection fields such that the mean of their absolute value across the field of view is 200 pixels, equal to 13\arcsec; see Appendix \ref{appendix:1} for more details). 

\begin{figure*}
\begin{center}
\begin{tabular}{c c c}
\includegraphics[width=56mm, height=54mm, trim= 70mm 10mm 80mm 10mm,clip]{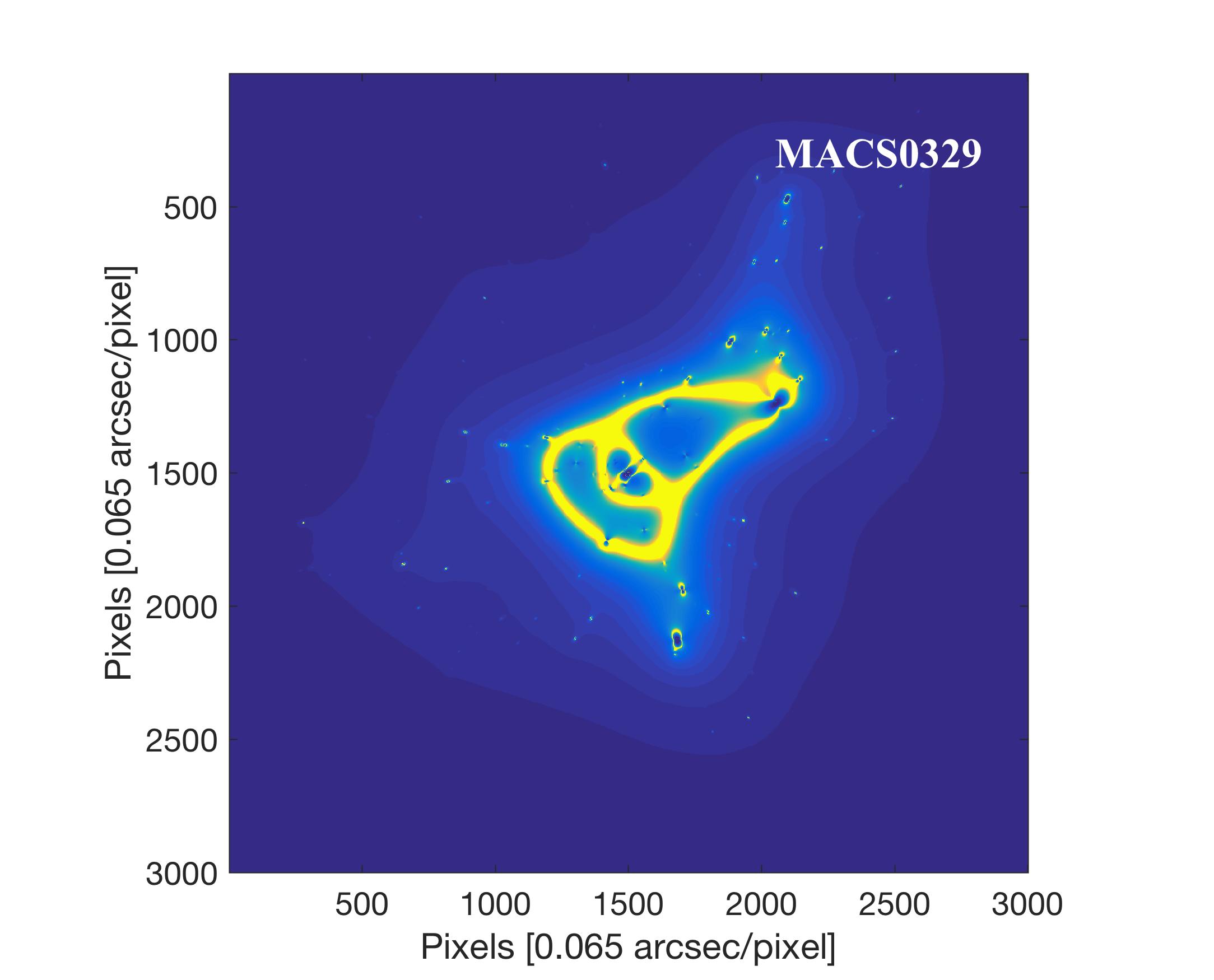} &
\includegraphics[width=56mm, height=54mm, trim= 70mm 10mm 80mm 10mm,clip]{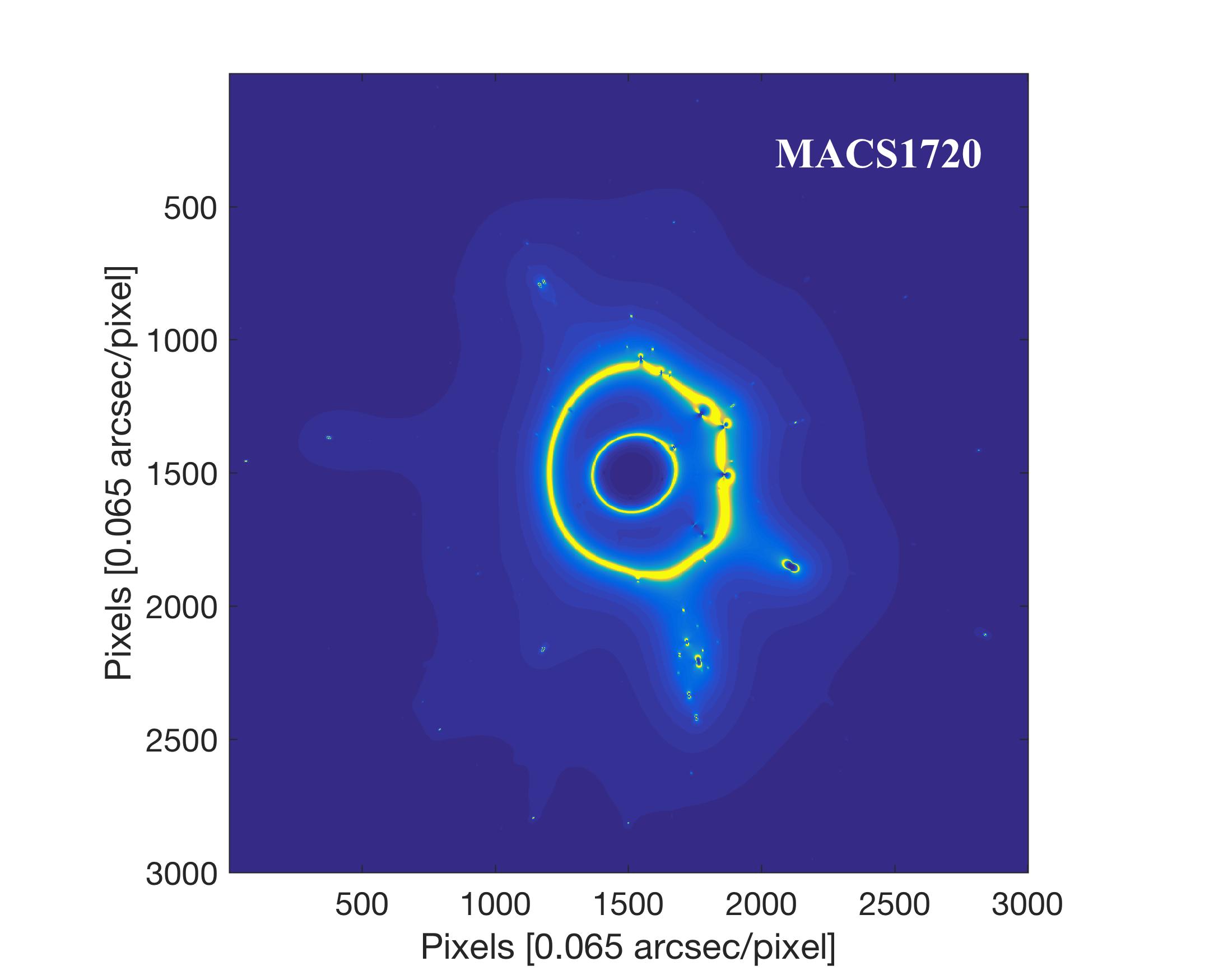} &
\includegraphics[width=56mm, height=54mm, trim= 70mm 10mm 80mm 10mm,clip]{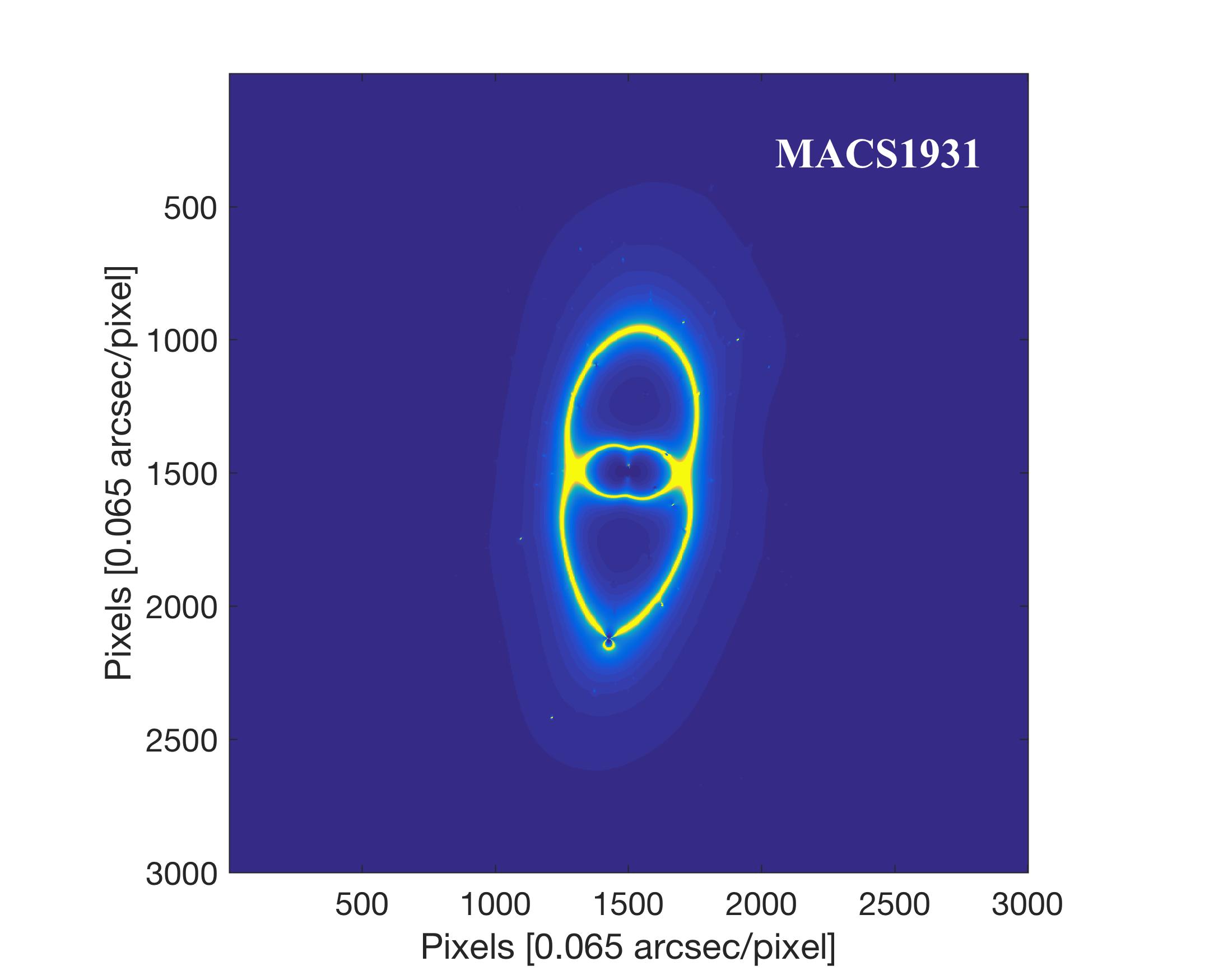} \\
\end{tabular}
\caption{\label{C9-fig:preliminary_mass_model}Magnification maps from the preliminary lens models for the clusters M0329 (left panel), M1720 (middle panel), and M1931 (right panel). 
}
\end{center}
\end{figure*}

As the input red-sequence list, including the BCG ellipticity parameters, we use the same member-galaxy catalogues used by \citet{Zitrin2015CLASH25}, based mainly on measurements with SExtractor \citep{BertinArnouts1996Sextractor} and a simple colour-magnitude red-sequence cut. In addition, we note that the flux measurements of BCGs, which are often much more diffuse than other cluster members, do not seem to trace well the encompassed mass of the BCGs (or, perhaps, BCGs may lie effectively on a different mass-to-light ratio compared to other galaxies, at least in our modelling scheme). We have found empirically in our previous modelling using this method, that the BCG $M/L$ ratio needs to be multiplied by a factor $\sim3-5$, which we adopt here as well.  We thus have now a single, preliminary mass model for each cluster (see Fig. \ref{C9-fig:preliminary_mass_model}) that can be used to find multiple images. It is worth mentioning that the construction of this initial mass model takes only a few seconds on a typical computer. Given it is key to the success of the automated multiple image finding, we make our code for constructing the initial lens-models publicly available\footnote{\color{blue}{https://github.com/adizitrin/initialLTM}}, and detail its implementation in Appendix \ref{appendix:1} below. Table \ref{modelparams} lists the parameter values used to construct each of the three models used here.


\subsection{Finding multiple images} \label{search_ima}
We describe now the simple algorithm used by the code for finding multiple images. Since the redshift, or $D_{ls}/D_s$ normalisation to which the preliminary mass model is scaled is not known a priori, to explore the relevant range the code probes 25 discrete values of $D_{ls}/D_s(z_s)$ normalisations, from source redshifts $z_s = z_l +0.3$ to the maximum source redshift given in the arc catalogue, where $z_l$ is fixed to the cluster redshift and $z_s$ increases such that the increments in $D_{ls}/D_s(z_s)$ are linear. In practice the 25 realisations are grouped in five principal $D_{ls}/D_s$ bins, each bin contains five sub-bins, corresponding to five different source redshifts, so that systems from the different sub-bins are merged together. The process now described refers to one $D_{ls}/D_s$ normalisation, and hence is repeated five times, once for each principal normalisation.

\subsubsection{Preliminary multiple image file}\label{tempimagefile}
The preliminary lens model is used to automatically project each candidate in the arc-candidate catalogue to the source plane and then back, to predict the location of possible counter images. This is done using the best-fit photometric redshift of the re-lensed arc as input. The program then runs over all the candidates in the catalogue searching for objects near the predicted locations and with similar photometric redshift and surface brightness as the re-lensed candidate. These will later be ranked according to their similarity to the re-lensed arc (\S \ref{ranking}). The process is repeated for the five sub-bins of the $D_{ls}/D_s$ normalisation adopted. To reduce spurious detections, thresholds are defined past which potential counter image candidates are not considered. The typical thresholds we probe are (in different iterations and for the different clusters -- see Table \ref{mifalparams} for the final parameters for each cluster) $\Delta z =0.2$ to $\Delta z =0.7$ for the redshift difference, $\Delta r = 5\arcsec - 10\arcsec$ for the search radius, and a relative $\Delta SB =20\%$ to $\Delta SB =300\%$, compared to the input re-lensed arc.

Once MIFAL has listed all the possible multiple images for the re-lensed candidate, for all candidates in the file and for the five normalisation sub-bins, it takes on a few organising steps -- such as removal of single images, combination of repeated images (i.e., if these correspond to the same arc), and coaddition of the lists from the five different sub-bins -- towards creating a preliminary list of potential multiple image systems for the adopted $D_{ls}/D_s$ normalisation. It then calculates the median redshift of each temporary system, $z_{sys}$, using the redshift of all members of that system, and registers the number of images in each system. At the end of this stage, each candidate in our catalogue has its own candidate multiple-image system and its respective $z_{sys}$. The file now is a preliminary list of the potential multiple image systems. Note that at this point this file may include some repetitions of systems, in the sense that if arc $i$ was relensed, and arcs $j$ and $k$ are listed as possible counter images so that these make a multiple image system, then the same system may appear again when listing candidate counter images for arcs $j$, and $k$. Also possible is the repetition of systems that are reproduced in more than one normalisation sub-bin.  

\subsubsection{The semifinal list}\label{reorganization}
The program now makes a few more reorganising steps to the preliminary multiple image list. First, it assigns a temporary index to each system. The index assignment is made following a simple count of the different systems, with the exclusion that systems originating from the same relensed arc, and that have similar median redshifts ($< \Delta z$),  are identified as the same system and given the same system number.  
This allows for grouping the same systems together (from different sub-bins, for example), which is the second organising step. 
Then, the code merges images that repeat in the same system, averaging their properties where differences exist (for example, the $D_{ls}/D_s$ may differ if images originate in different sub-bins), and recalculates the median redshift of each system. 

The next stage is to obtain the so-called semifinal list of multiple images (per adopted $D_{ls}/D_s$ normalisation), merging together systems that have at least 2 images in common and similar redshift ($z_{sys} \pm \Delta z$), 
merging again images that repeat in each new system, and recalculating the system's properties such as median redshift and number of images. 

\subsubsection{Final multiple image lists}\label{ranking}

To obtain a final list of multiple images, MIFAL now assigns to each multiple-image candidate a $\chi^2$ grade to estimate its likelihood of being a member of a certain family. The $\chi^2$ for each image is defined by:

\begin{equation}\label{chi2_ima_1}
\chi^{2}_{ima}=\chi^{2}_{r}+\chi^{2}_{z}+\chi^{2}_{colour}+\chi^{2}_{SB},
\end{equation}

\noindent where the first component is the $\chi^{2}$ of the image location with respect to the expected position. This is defined by:

\begin{equation} \label{chi2_r}
\chi^{2}_{r}=((x'-x)^2 + (y'-y)^2) ~/ ~\sigma^{2}_{r},
\end{equation}

\noindent with [x$'$,y$'$] and [x,y] being the position of the candidate, and the model-predicted multiple-image, respectively, and $\sigma_{r}$ the expected image-plane
reproduction uncertainty. Motivated by studies that probed the contribution from large-scale structure and matter along the line of sight to the lensing signal \citep[e.g.][]{Host2012LOS}, we set $\sigma_{r} = 1.4''$  throughout. The second component is defined as:

\begin{equation} \label{chi2_z}
\chi^{2}_{z}=((z'-\overline{z})^2) ~/ ~\sigma^{2}_{z},
\end{equation}

\noindent where $z'$, $ \overline{z}$, and $\sigma^{2}_{z}$, represent the measured photometric redshift, the median system redshift, and the dispersion of photometric redshift measurements within the system, respectively. The third component is defined as:

\begin{equation} \label{chi2_c}
\chi^{2}_{colour}=\sum_{i} (colour'_{i}-\overline{colour}_{i})^2 ~/ ~\sigma^{2}_{colour_i},
\end{equation}

\noindent where, $colour'_{i}$, $\overline{colour}_{i}$, and $\sigma^{2}_{colour_i}$ are the $i'th$ colour of the candidate in question, the median value of this colour in the candidate system (from the other images found for that system excluding the image in question), and its dispersion, respectively. While any combination of colours can be chosen, and while CLASH has more filters available, in practice we limit our case study to using only the following colours (over which the sum is performed): F475W -- F435W, F625W -- F475W, F814W -- F625W, F814W -- F475W, F105W -- F814W, F125W -- F105W, and F140W -- F125W. If there are non detections in some bands, or if the candidate arc is a dropout, i.e., a Lyman break galaxy, for example, then the relevant colours are ignored. The fourth component is defined as:

\begin{equation} \label{chi2_SB}
\chi^{2}_{SB}=((SB' - SB )^2 ) ~/ ~\sigma^{2}_{SB},
\end{equation}

\noindent where $SB'$ is a surface brightness measure of the candidate that was lensed (in practice we use the F814W flux with a fixed, arbitrary zero point, divided by the area used for the flux measurement of each candidate). $SB$ is the surface brightness measure of the candidate counter image, and $\sigma^2_{SB}$  the respective $1-\sigma$ error given by the dispersion of surface brightness measures within the system.  

After MIFAL assigns a $\chi^2$ grade to each candidate, it selects the so called final, most likely images per system that are below a user-defined $\chi^2_{ima}$ threshold (which we take throughout as 50). In practice, the allowed number of images in a system is typically being limited to at least 2 (to minimise spurious detections, a  minimum of 3 images can be used). The maximum number of images in a system we use is 4. Thus, if the semifinal system has between two and four images below the threshold, the program now adopts it as is. If it has five images or more (some of the semifinal systems may have them), the algorithm selects the four images with the lowest $\chi^2_{ima}$ that meet the predefined $\chi^2_{ima}$ threshold. 
Then, it calculates the total $\chi^2_{sys}$ of the system, simply adding the $\chi^2_{ima}$ of the selected images. For now, we allow also for multiple images to be considered in more than one system (with different $\chi^2_{ima}$ weights), although this can clearly be modified to choose only a single system that corresponds to the best $\chi^2$. After each system has been finalised, the program recalculates again its properties such as median redshift and updated number of images.

\subsubsection{Repetition and normalisation}\label{repetition}
Once the program has run through all semifinal systems, filtering and updating them using the said $\chi^2$, we obtain the final multiple-image systems catalogue. Since, as mentioned, the code repeats the process described above for 5 principal $D_{ls}/D_s(z_s)$ normalisation bins, we end up with one multiple image catalogue for each $D_{ls} / D_s$ bin. This is the final stage of the current process. However, since multiple images have now been found these can be used to refine the model (e.g., using a quick Monte-Carlo Markov Chain (MCMC) minimisation), hence calibrating it to a specific chosen redshift. This calibrated model could now, in principle (i.e., we do not perform this additional step here) be used to repeat the MIFAL process, potentially refining the list of multiple images.

\subsubsection{Adding lens configuration and parity information}\label{adding}
The simple $\chi^2$ ranking uses only the metric defined above for grading each multiple image and system: the similarity in terms of colours, redshift, position, and surface brightness. In addition, typical lensing configurations, as well as the parity of the images, can be used to improve further the selection of multiple image families. For completeness, we briefly mention this possibility (partly incorporated in the code at present), although we do not present results based on these features in the current work, which is aimed merely to constitute a first proof-of-concept for the automated finding of multiple images.

If one assumes that most clusters would show, roughly, lensing configurations as those given by smooth elliptical mass
distributions, extra information can be incorporated to improve the multiple-image list: for example, to exclude two nearby images that are on the same side of the critical curve (and so the method could choose the preferred one based on its $\chi^{2}$ ranking). Some of the most typical lens configurations are the Einstein cross, cusp arc, and fold arc systems, which have five multiple-images each (four distributed around the lens and one, often smaller and fainter than the rest, close to the cluster centre). As is common in massive elongated clusters, cusp like systems often show only three highly magnified close images; a configuration that could be considered independently, as well. Hence the method can use these four lens configurations to add position constraints in the selection process of the final multiple images. 
In practice, this means the program would take each semifinal system and check if there are image combinations that match with some of these four lens
configurations, including the parity, and save or prioritise the successful image combinations. The program would assign a
$\chi^2$ to each successful combination, that is combined with the previous $\chi^2$ ranking.

\begin{figure*}
\begin{center}
\begin{tabular}{c}
\includegraphics[width=120mm, trim= 0mm 0mm 0mm 0mm,clip]{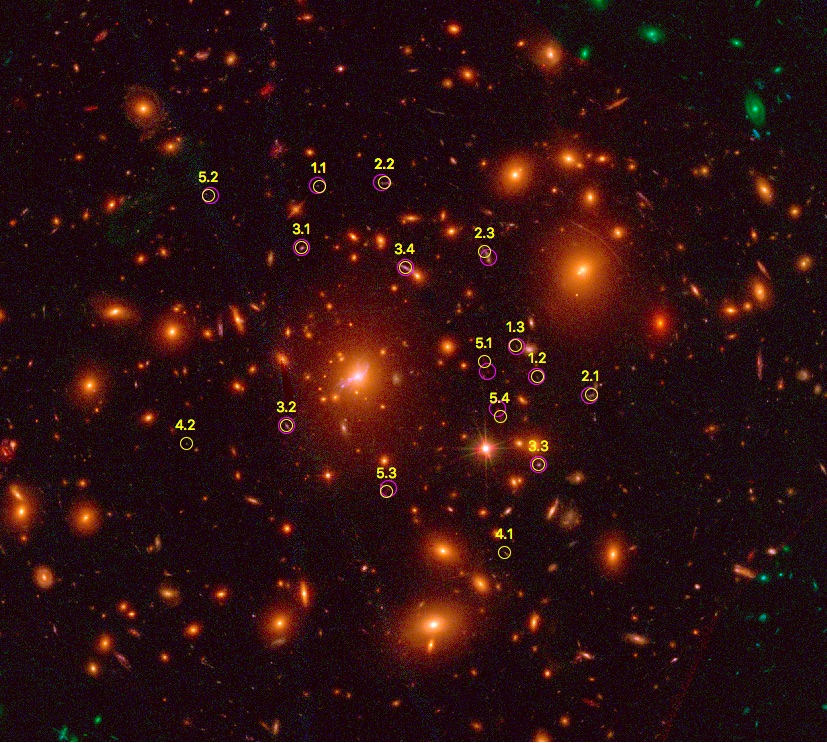} \\
\end{tabular}
\caption{\label{C9-fig:macs0329}
CLASH RGB image of M0329 with the most likely multiple-image systems found by MIFAL labelled by \emph{yellow circles}, for 
a normalisation redshift of $z_{norm} =1.46$. \emph{Magenta circles} show the systems previously identified \citep{Zitrin2015CLASH25}, for comparison. Hence yellow images with no magenta counterparts are either new potential systems or false detections, and magenta circles with no yellow counterparts are known images that have not been recovered. Systems 2, 3, and 5 seen in the Figure as recovered by MIFAL are known identifications from \citet{Zitrin2015CLASH25}. System 1 (and images 1.2 \& 1.3 in particular) is a potential (less secure) candidate system previously identified, and system 4 is a false detection.
}
\end{center}
\end{figure*}
\begin{table*}
\caption{MIFAL results for M0329\label{C9-table:m0329}}
\centering
\resizebox{2.1\columnwidth}{!}{
\begin{tabular}{c c c c c c c c c c c c c c}
\toprule \toprule
Id & R.A. &   Dec &   $z_{ima}$ &  $z_{sys}$ & $z_{norm}$ & $R_{find}$ & $\chi^2_{col}$ &  $\chi^2_{z}$ & $\chi^2_{R_{find}} $ & $\chi^2_{SB}$ & $\chi^2_{ima}$ & $\chi^2_{sys}$ & Comment\\ 
& $(^{\circ})$  & $(^{\circ})$   &  &   &  & $('')$ & & &  &  &  & & \\ 
\midrule 
1.1 & 52.424900 & -2.187690 & 3.25  & 3.26 &  1.46 &  2.3 &  0.60 &  0.00 &  2.74 &  2.30 &  5.63 &  23.74 & Sys. c5 in Z15\\ 
1.2 & 52.415100 & -2.196250 & 3.26  & 3.26 &  1.46 &  2.3 &  13.08 &  0.00 &  2.69 &  0.00 &  15.78 &  23.74 & "\\ 
1.3 & 52.416100 & -2.194830 & 3.26  & 3.26 &  1.46 &  0.4 &  2.24 &  0.00 &  0.08 &  0.00 &  2.32 &  23.74 &"\\ 
2.1 & 52.412700 & -2.197070 & 2.87  & 2.83 &  1.46 &  1.6 &  2.16 &  0.02 &  1.33 &  1.71 &  5.23 &  40.05 & Sys. 3 in Z15\\ 
2.2 & 52.422000 & -2.187520 & 2.79  & 2.83 &  1.46 &  1.9 &  17.78 &  0.11 &  1.92 &  8.93 &  28.74 &  40.05 &"\\ 
2.3 & 52.417500 & -2.190630 & 2.83  & 2.83 &  1.46 &  2.3 &  2.91 &  0.03 &  2.67 &  0.47 &  6.07 &  40.05 &"\\ 
3.1 & 52.425725 & -2.190436 & 2.13  & 2.19 &  1.46 &  3.0 &  0.23 &  0.34 &  4.49 &  0.25 &  5.31 &  34.48 & Sys. 2 in Z15\\ 
3.2 & 52.426393 & -2.198453 & 2.24  & 2.19 &  1.46 &  2.3 &  3.83 &  1.76 &  2.66 &  0.19 &  8.44 &  34.48 &"\\ 
3.3 & 52.415078 & -2.200176 & 2.26  & 2.19 &  1.46 &  2.7 &  1.87 &  3.02 &  3.66 &  1.12 &  9.68 &  34.48 &"\\ 
3.4 & 52.421065 & -2.191334 & 2.13  & 2.19 &  1.46 &  1.8 &  1.65 &  0.34 &  1.65 &  7.41 &  11.05 &  34.48 &"\\ 
4.1 & 52.416600 & -2.204170 & 3.20  & 3.06 &  1.46 &  4.7 &  14.00 &  3.02 &  11.39 &  16.13 &  44.54 &  51.35 & False detection\\ 
4.2 & 52.430903 & -2.199237 & 2.93  & 3.06 &  1.46 &  0.2 &  5.06 &  0.01 &  0.02 &  1.72 &  6.81 &  51.35 & "\\ 
5.1 & 52.417500 & -2.195560 & 5.79  & 6.06 &  1.46 &  1.9 &  1.97 &  4.16 &  1.89 &  10.41 &  18.42 &  48.62 &Sys. 1 in Z15\\ 
5.2 & 52.429900 & -2.188120 & 6.08  & 6.06 &  1.46 &  3.9 &  0.66 &  0.01 &  7.87 &  1.43 &  9.98 &  48.62 &"\\ 
5.3 & 52.421900 & -2.201400 & 6.05  & 6.06 &  1.46 &  2.5 &  0.89 &  0.01 &  3.12 &  0.47 &  4.49 &  48.62 &"\\ 
5.4 & 52.416800 & -2.198060 & 6.31  & 6.06 &  1.46 &  4.9 &  0.14 &  2.66 &  12.39 &  0.55 &  15.73 &  48.62 &"\\ 
\bottomrule
\end{tabular}
}                                        

{\footnotesize\flushleft
Note - MIFAL-identified multiple image systems for a normalisation of $z_{s}=2.05$ and thresholds of [$\Delta z =0.7$, $\Delta r = 6.5\arcsec$, relative $\Delta SB =50\%$]. See \S \ref{SLanalysis} for more details. Z15 stands for \citep{Zitrin2015CLASH25}.\\
$Column~1:$ image Id; $Columns~2~\&~3:$ coordinates in J2000; Cluster name and redshift; $Column~4:$ photometric redshift of the image; $Column~4:$ median photometric redshift as the redshift for the system; $Column~6:$ source redshift to which the initial model is normalised; $Column~7:$ average distance of the image from its predicted location; $Column~8-11:$ $\chi^2$ estimate of the arc to belong to the system based on colour, redshift, distance from predicted location, and surface brightness, respectively; $Column~12:$ the total $\chi^2$ of the system; $Column~13:$ comments: mainly for designating whether the automatically identified image is likely real or not.\\
}
\end{table*}

\section{Results}\label{SLanalysis}

As the first proof-of-concept for the presented method and the feasibility for detecting multiple images in a ``blind" fashion, we analyse three CLASH clusters: M0329, M1720, and MACS1931. These clusters were previously analysed in the framework of the CLASH program by \citet{Zitrin2015CLASH25}, who found several multiple imaged systems in each of them (see also \citealt{Caminha2019_clash}). The first two clusters, M0329 and M1720, show a complex, unrelaxed, or substructured central mass distribution, whereas the last one, M1931, shows quite an elongated, elliptical-like shape.

As detailed in \S \ref{algo}, we run the Arcfinder on all three cluster detection images, and then cross-match the arcfinder's results to the CLASH catalogues to obtain the photometry and photometric redshift of each object. We then clean the catalogue from objects lying outside the model's FOV, or that do not lie far enough in redshift beyond the cluster's redshift. We construct preliminary mass models for each cluster, as detailed in \S \ref{pre_mass_model}. The preliminary, ``blind" mass models of these clusters, constructed without any input regarding the multiple images, are shown in terms of their magnification maps in Fig. \ref{C9-fig:preliminary_mass_model}, and the parameters used for constructing these are given in Appendix \ref{appendix:2}. We run MIFAL on these three clusters, and detail here the results. To probe the method with respect to the input thresholds, we start with the same thresholds for all three clusters [$\Delta z =0.5$, $\Delta r = 6.5\arcsec$, relative $\Delta SB =50\%$] and slightly refine them per cluster, to probe the effect on the results. The most likely multiple-image systems found by the automated procedure are reported in Tables \ref{C9-table:m0329}, \ref{C9-table:m1720}, and \ref{C9-table:m1931}. 

\noindent\textbf{MACS J0329.6-0211:}  \citet[see also][]{Zitrin2012CLASH0329,Zitrin2015CLASH25} reported six sets of likely lensed systems in M0329, although only three systems they considered secure and were used to constrain their model. For comparison, Fig. \ref{C9-fig:macs0329} shows the multiple-image systems found by the automated MIFAL for a normalisation redshift of $z_{norm} =1.46$, and the following thresholds [$\Delta z =0.5$, $\Delta r = 5.2\arcsec$, relative $\Delta SB =300\%$]. Without any multiple image constraints as input, MIFAL fully recovers the three multiple-image systems, including the quadruply imaged $z\sim6$ dropout galaxy. Another candidate system proposed by \citet{Zitrin2015CLASH25} is also fully recovered. In addition, there is a seemingly false two-image system also proposed (system 4 here), which can be discarded by either accepting only systems that have three and above multiple images, or by applying thresholds to the different $\chi^2$'s. Note also that with the default thresholds MIFAL initially suggested two other (most likely unreal) systems, that were either excluded by the code based on their very high $\chi^2$ -- that is 1-2 orders of magnitude higher than the secure systems -- or were not detected when revising the thresholds as above. Especially, lowering $\Delta r$ to  $5.2\arcsec$ excluded some other suggested counter image candidates. We also note that when $\Delta SB$ was set initially to $50\%$, the $z\sim6$ dropout system went missing (because its F814W flux is noisy) -  but was recovered within the relative $\Delta SB =50\%$ threshold if we instead calculated the SB based on a redder band (F125W). In that case all MIFAL systems seen in Fig. \ref{C9-fig:macs0329} are discovered, aside from image 2.1, and with no spurious detections. We thus conclude that of the probed combinations, that is the most successful set of thresholds for M0329. The systems are also listed in Table \ref{C9-table:m0329}. These are sufficient to fully constrain a proper lens model for the cluster.

\begin{figure*} 
\begin{center}
\begin{tabular}{c}
\includegraphics[width=120mm, trim= 0mm 0mm 0mm 0mm,clip]{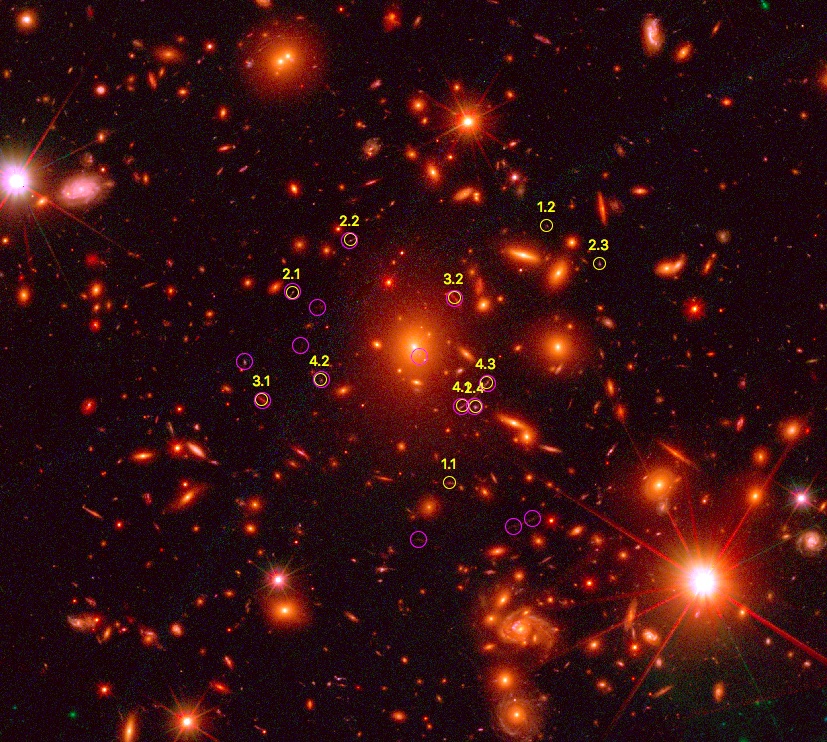} \\
\end{tabular}
\caption{\label{C9-fig:macs1720}
CLASH RGB image of M1720 with the most likely multiple-image systems found by MIFAL labelled by \emph{yellow circles}, for 
a normalisation redshift of $z_{norm} =3.85$. \emph{Magenta circles} show the systems previously identified \citep{Zitrin2015CLASH25}, for comparison. Hence yellow images with no magenta counterparts are either new potential systems or false detections, and magenta circles with no yellow counterparts are known images that have not been recovered. Systems 2, 3, and 4 seen in the Figure as recovered by MIFAL are known identifications from \citet{Zitrin2015CLASH25}, with 2.3 a likely new match for the system and 2.4 being less certain. System 1 is most likely a false detection.
}
\end{center}
\end{figure*}

\begin{table*}

\caption{MIFAL results for M1720\label{C9-table:m1720}}

\centering
\resizebox{2.1\columnwidth}{!}{
\begin{tabular}{c c c c c c c c c c c c c c}
\toprule \toprule

Id & R.A.  &   Dec &   $z_{ima}$ &  $z_{sys}$ & $z_{norm}$ & $R_{find}$ & $\chi^2_{col}$ &  $\chi^2_{z}$ & $\chi^2_{R_{find}} $ & $\chi^2_{SB}$ & $\chi^2_{ima}$ & $\chi^2_{sys}$ & Comment\\ 
& $(^{\circ})$  & $(^{\circ})$   &  &   &  & $('')$ & & &  &  &  & & \\ 
\midrule 
1.1 & 260.067820 & 35.601192 & 2.47  & 2.39 &  3.85 &  5.9 &  22.94 &  0.65 &  17.69 &  0.39 &  41.67 &  65.64 & False identification\\ 
1.2 & 260.062460 & 35.612745 & 2.31  & 2.39 &  3.85 &  0.2 &  22.94 &  0.38 &  0.02 &  0.63 &  23.97 &  65.64 & "\\ 
2.1 & 260.076530 & 35.609749 & 3.22  & 2.88 &  3.85 &  3.0 &  0.47 &  0.85 &  4.51 &  0.18 &  6.01 &  79.88 &Sys. 1 in Z15\\ 
2.2 & 260.073320 & 35.612099 & 3.11  & 2.88 &  3.85 &  4.4 &  1.17 &  0.16 &  10.06 &  0.10 &  11.49 &  79.88 &"\\ 
2.3 & 260.059560 & 35.611026 & 2.66  & 2.88 &  3.85 &  5.7 &  3.06 &  3.54 &  16.85 &  0.31 &  23.75 &  79.88 &Likely\\ 
2.4 & 260.066420 & 35.604602 & 2.65  & 2.88 &  3.85 &  7.7 &  4.68 &  3.79 &  30.07 &  0.10 &  38.64 &  79.88 & Uncertain\\ 
3.1 & 260.078200 & 35.604918 & 1.68  & 1.68 &  3.85 &  7.7 &  12.69 &  0.50 &  30.26 &  3.68 &  47.13 &  50.84 &Sys. 3 in Z15\\ 
3.2 & 260.067580 & 35.609483 & 1.68  & 1.68 &  3.85 &  0.2 &  3.18 &  0.50 &  0.02 &  0.00 &  3.70 &  50.84 &"\\ 
4.1 & 260.067140 & 35.604648 & 0.98  & 0.86 &  3.85 &  2.5 &  2.51 &  14.06 &  3.11 &  1.66 &  21.35 &  71.33 &Sys. 2 in Z15\\ 
4.2 & 260.074950 & 35.605803 & 0.84  & 0.86 &  3.85 &  5.3 &  23.21 &  0.05 &  14.48 &  5.31 &  43.04 &  71.33 &"\\ 
4.3 & 260.065810 & 35.605677 & 0.86  & 0.86 &  3.85 &  2.7 &  1.10 &  0.04 &  3.77 &  2.03 &  6.94 &  71.33 &"\\ 

\bottomrule
\end{tabular}
}                                        

{\footnotesize\flushleft
Note - MIFAL-identified multiple image systems for a normalisation of $z_{s}=2.05$ and thresholds of [$\Delta z =0.7$, $\Delta r = 6.5\arcsec$, relative $\Delta SB =50\%$]. See \S \ref{SLanalysis} for more details. \\
$Column~1:$ image Id; $Columns~2~\&~3:$ coordinates in J2000; Cluster name and redshift; $Column~4:$ photometric redshift of the image; $Column~4:$ median photometric redshift as the redshift for the system; $Column~6:$ source redshift to which the initial model is normalised; $Column~7:$ average distance of the image from its predicted location; $Column~8-11:$ $\chi^2$ estimate of the arc to belong to the system based on colour, redshift, distance from predicted location, and surface brightness, respectively; $Column~12:$ the total $\chi^2$ of the system; $Column~13:$ comments: mainly for designating whether the automatically identified image is likely real or not.\\
}

\end{table*}

\noindent\textbf{ MACS J1720+3536:}  \citet{Zitrin2015CLASH25} reported 5 secure systems in M1720, and one additional candidate system (system 5 in their notation).  Fig. \ref{C9-fig:macs1720} shows the most likely multiple-image systems found by MIFAL for a normalisation redshift of $z_{norm} =3.85$, and the following thresholds [$\Delta z =0.5$, $\Delta r = 7.8\arcsec$, relative $\Delta SB =300\%$]. Our automated algorithm recovers three of the five secure systems by \citet{Zitrin2015CLASH25}, including a distinct, bright two-image red system. The two remaining systems from \citet{Zitrin2015CLASH25} we do not expect to be recovered by MIFAL, since many of their images are not present in the automatic arc detection+photometric redshift catalogue. Among the three systems that are recovered by MIFAL, two (systems 3 \& 4 here) are exactly reproduced, while for system 2 three images seem to be well matched but an additional image is also included (image 2.4), which may not be related. Additionally, \citet{Zitrin2015CLASH25} considered another possible counter image as image 2.3, but MIFAL's choice seems reasonable. The proposed two-image system number 1, seems to be falsely detected altogether. If we use the default thresholds [$\Delta z =0.5$, $\Delta r = 6.5\arcsec$, relative $\Delta SB =50\%$], then only system 2 and false system 1 are recovered. Both the SB threshold and the radius have to be increased to detect the remaining systems. The multiple-image systems and their properties, for the chosen thresholds, can be found in Table \ref{C9-table:m1720}. Therefore, in this case our algorithm was able to recover  $\sim2.5$ out of 5 previously detected systems found by aid of a visual inspection. Still, the procedure supplies enough secure constraints to revise the initial lens model.

\begin{figure*} 
\begin{center}
\begin{tabular}{c}
\includegraphics[width=120mm, trim= 0mm 0mm 0mm 0mm,clip]{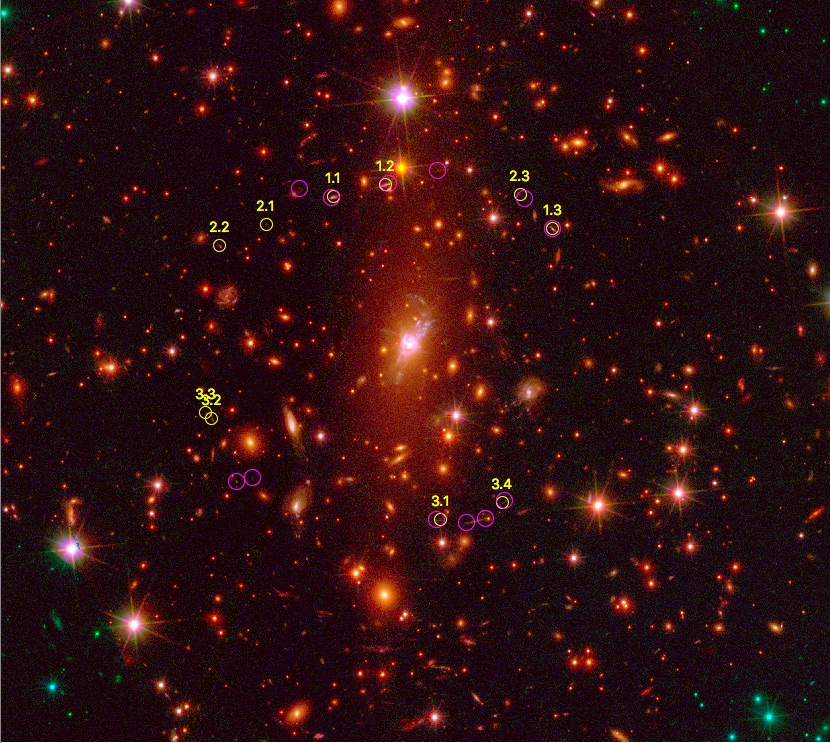} \\
\end{tabular}
\caption{\label{C9-fig:macs1931}
CLASH RGB image of M1931 with the most likely multiple-image systems found by MIFAL labelled by \emph{yellow circles}, for 
a normalisation redshift of $z_{norm} =2.03$. \emph{Magenta circles} show the systems previously identified \citep{Zitrin2015CLASH25}, for comparison. Hence yellow images with no magenta counterparts are either new potential systems or false detections, and magenta circles with no yellow counterparts are known images that have not been recovered. System 1 seen in the Figure as recovered by MIFAL is a known identification from \citet{Zitrin2015CLASH25}. Images 3.1 \& 3.4 also belong to a known system but image 3.2 differs from the nearby identification by \citet{Zitrin2015CLASH25}, and a fourth image (3.3) is being listed as well but is most likely unrelated. System 2 is erroneously identified.
}
\end{center}
\end{figure*}

\noindent\textbf{MACS J1931-2635:} \citet{Zitrin2015CLASH25} reported 4 secure systems in M1931 (and three other potential candidate systems, that we disregard). Fig. \ref{C9-fig:macs1931} shows the most likely multiple-image systems found by MIFAL for a normalisation redshift of $z_{norm} =2.03$, and the following thresholds [$\Delta z =0.7$, $\Delta r = 6.5\arcsec$, relative $\Delta SB =50\%$]. It is first noted that MIFAL recovers well system 1. System 1 shows a large scatter in photometric redshifts, probably due to contaminated photometry by nearby stars or cluster members, and so the initial default $\Delta z$ had to be raised to $0.7$ to recover it. MIFAL also recovers a second system, namely two images of system 3 here (3.1 \& 3.4; images 4.1 \& 4.2 in the notation of \citealt{Zitrin2015CLASH25}). As the counter image for this system expected on the other side of the cluster MIFAL assigns a different image than the one \citet{Zitrin2015CLASH25} found, but appears to be similar-looking as well. In fact MIFAL lists two nearby images as the counter image but if these are not a single object, only one of them can be the true counter image. Next, we do not expect MIFAL to reproduce system 2 from \citet{Zitrin2015CLASH25} since two of its images do not appear in the arc+photometry catalogue, and indeed MIFAL assigns two other (false) counter images to image 2.3 (image 2.3 also in their notation), which is the only one from this system that appears in the catalogue. Similarly, we do not expect MIFAL to reproduce system 3 from \citet{Zitrin2015CLASH25}, because it is only partly listed in the arc+photometry catalogue and hence also absent from the final multiple-image catalogue.  The multiple image systems are listed in Table \ref{C9-table:m1931}. Overall MIFAL does well in reproducing the systems that appear in the catalogue for this cluster, with some differences in the image designated as 3.3.

\begin{table*}
\caption{MIFAL results for M1931\label{C9-table:m1931}}

\centering
\resizebox{2.1\columnwidth}{!}{
\begin{tabular}{c c c c c c c c c c c c c c c c}
\toprule \toprule

Id & R.A.  &   Dec &   $z_{ima}$ &  $z_{sys}$ & $z_{norm}$ & $R_{find}$ & $\chi^2_{col}$ &  $\chi^2_{z}$ & $\chi^2_{R_{find}} $ & $\chi^2_{SB}$ & $\chi^2_{ima}$ & $\chi^2_{sys}$ & Comment\\ 
& $(^{\circ})$  & $(^{\circ})$   &  &   &  & $('')$ & & &  &  &  & & \\ 
\midrule 
1.1 & 292.960600 & -26.569164 & 1.80  & 1.80 &  2.03 &  1.1 &  1.31 &  0.03 &  0.58 &  0.01 &  1.94 &  32.90 &Sys. 1 in Z15\\ 
1.2 & 292.957990 & -26.568611 & 1.41  & 1.80 &  2.03 &  0.8 &  2.44 &  2.35 &  0.32 &  2.99 &  8.10 &  32.90 &"\\ 
1.3 & 292.949590 & -26.570605 & 2.46  & 1.80 &  2.03 &  3.3 &  0.53 &  9.78 &  5.45 &  7.09 &  22.86 &  32.90 &"\\ 
2.1 & 292.963980 & -26.570404 & 2.22  & 2.23 &  2.03 &  3.9 &  20.38 &  0.00 &  7.70 &  0.87 &  28.95 &  72.29 &False identification\\ 
2.2 & 292.966320 & -26.571364 & 2.28  & 2.23 &  2.03 &  5.1 &  6.55 &  0.03 &  13.22 &  16.48 &  36.28 &  72.29 &False identification\\ 
2.3 & 292.951220 & -26.569077 & 2.23  & 2.23 &  2.03 &  0.2 &  6.69 &  0.00 &  0.02 &  0.34 &  7.05 &  72.29 &Image 2.3 in Z15\\ 
3.1 & 292.955220 & -26.583679 & 2.90  & 2.92 &  2.03 &  2.3 &  5.10 &  0.09 &  2.65 &  1.93 &  9.76 &  74.10 &Sys. 4 in Z15\\ 
3.2 & 292.966700 & -26.579142 & 3.06  & 2.92 &  2.03 &  3.7 &  2.88 &  0.86 &  6.85 &  0.17 &  10.75 &  74.10 &Differs from Z15 but possible\\ 
3.3 & 292.967010 & -26.578874 & 2.81  & 2.92 &  2.03 &  5.7 &  7.92 &  0.08 &  16.85 &  0.17 &  25.02 &  74.10 &Nearby but likely unrelated\\ 
3.4 & 292.952110 & -26.582897 & 2.95  & 2.92 &  2.03 &  2.7 &  22.12 &  0.22 &  3.72 &  2.51 &  28.57 &  74.10 &Sys. 4 in Z15 \\

\bottomrule
\end{tabular}
}                                        

{\footnotesize\flushleft
Note - MIFAL-identified multiple image systems for a normalisation of $z_{s}=2.05$ and thresholds of [$\Delta z =0.7$, $\Delta r = 6.5\arcsec$, relative $\Delta SB =50\%$]. See \S \ref{SLanalysis} for more details. \\
$Column~1:$ image Id; $Columns~2~\&~3:$ coordinates in J2000; Cluster name and redshift; $Column~4:$ photometric redshift of the image; $Column~4:$ median photometric redshift as the redshift for the system; $Column~6:$ source redshift to which the initial model is normalised; $Column~7:$ average distance of the image from its predicted location; $Column~8-11:$ $\chi^2$ estimate of the arc to belong to the system based on colour, redshift, distance from predicted location, and surface brightness, respectively; $Column~12:$ the total $\chi^2$ of the system; $Column~13:$ comments: mainly for designating whether the automatically identified image is likely real or not.\\
}

\end{table*}

\section{Discussion}\label{Discussion}
We have probed the feasibility of finding multiple images automatically, on three massive clusters imaged with the HST. The outlined procedure was operated on the three clusters and compared to previous analyses. The method seems to be quite successful: without any multiple images used as input, it recovers all secure known multiple images in M0329, and several secure multiple-image systems in M1720, and M1931. Given the trade off between completeness of true images and spurious detections, some erroneous multiple images were identified as well, and thresholds had to be refined to maximise the true-to-false identification ratio. The number of identified secure systems is sufficient to fully constrain, or refine, the initial lens models. In fact, the output multiple image list can be used as input for any lens modelling technique and so the final solution does not necessarily have to be limited to the LTM technique.

The method we presented for automatically finding multiple images consists of three components: the arc candidate catalogue, a photometry and photometric redshift catalogue, and the initial lens model.  Each of these constitute a crucial ingredient for the success of the method. The arc-finder and the photometry are known to operate very well, yielding stable results. The promise of the process is perhaps driven by the initial lens model -- such blind automated lens modeling for SL clusters are not yet too common -- constructed following a revised version of the LTM methodology \citep[][see details of the LTM code in Appendix \ref{appendix:1}]{Zitrin2009b,Zitrin2015CLASH25}. 

The LTM is very promising for the automated analysis of SL clusters, given that the initially guessed model, based only on the distribution of cluster members and their luminosity, is already reasonable enough so that multiple images can be found using that model (see as a few examples \citealt{Zitrin2009b,Zitrin2015CLASH25}, and references therein). However, the LTM initial-guess model, even while evidently successful, cannot be expected to trace the mass distribution exactly, and hence its success could differ from cluster to cluster. In addition, compared to analytic models such as elliptical or pseudo elliptical mass distributions, the LTM methodology has a few limitations. First, the mass distribution is often rounder, where the required ellipticity of the critical curves often needed, is introduced as an external shear contribution. This results in a limited overall elongation of the model. Second, the mass profile is often shallower, which results in higher magnifications compared to common analytical models. Finally, it is susceptible to galaxies that lie outside the adopted M/L ratio, which can significantly bias the resulting model. 

This range of success levels is only partly manifested in the analysis we perform here, where the initial model for the three clusters seems to be successful in finding all or most multiple images using the automated procedure, at least, those that are included in the arc-photometry catalogue. MIFAL has a harder time reproducing some images but this, however, seems to be in large part a more direct result of noisy or contaminated photometry, and erroneous photometric redshifts, in some cases.

The current work aims to constitute a proof-of-concept for the automated finding of multiple images. Clearly, there are many avenues for improving and further scrutinising the presented procedure. Such improvements, for example, include the use of additional information such as known lensing configurations, parity and/or shear or magnification of the images, using the full photometric-redshift probability distribution of each arc (rather then just the best-fit value), or explicit use of the structure of the images (i.e., internal details) through a full-arc reconstruction comparison for each image. In addition, ways to improve the initial model guess, including the usage of other alternative parametrisations, should be probed as well. Similarly, one major improvement could be to find the effective scaling relation between the luminosity and mass that corresponds to the current methodology, as was done, for example, by \citep{Zitrin2011d}. It might also be beneficial to examine whether the weighting of the different $\chi^2$'s should remain as presented or perhaps modified, to give more weight to certain, more stable measurements. 

While the success of the method is promising, there is now a need to continue operating the method on more clusters, to continue examining its performance in different configurations. 

Finally, it should also be mentioned that the current procedure is one example of many possible procedures, including, conceivably, those that combine machine learning algorithms for image identification.

\section{Summary}\label{summary}
In this work we presented an innovative \textbf{M}ultiple-\textbf{I}mage \textbf{F}inder \textbf{AL}gorithm (MIFAL)
designed to automatically find multiple images in galaxy cluster lenses, so that
their mass model could be efficiently and automatically constrained. We combine an \emph{arcfinder} algorithm with CLASH photometric redshift measurements,
along with a preliminary mass model, to physically match together multiple-image systems in an automated (``blind'') manner.
We obtain a robust assessment of the likelihood of each arc to belong to one of the multiple-image systems, as well as the preferred redshift for the
different systems. MIFAL then selects the most likely multiple images for each system based on the assigned grades, to finally construct a catalogue of multiple-image systems.

We applied MIFAL to the recently studied galaxy clusters M0329, M1720 and M1931 in deep CLASH/HST images.
We compared the results of our automated procedure with the results by the conventional SL analysis where multiple images were verified
by eye and using parametric lens reconstruction methods. Our automated SL analysis by MIFAL blindly recovered all three known systems in M0329, and about two or three, or half of the systems in M1720 and M1931. In cases where multiple images and/or systems were not picked by the automated procedure, this was mainly due the arcs being faint and missing from the original arc catalogue. Some false identification also occurred, mostly due to ambiguous or erroneous photometric redshift measurements, or in part, inaccurate initial lens model.

Although more comparisons are required for a more robust assessment of the proposed algorithm accuracy, and although the basic proof-of-concept procedure we outline here is perhaps not yet fully competitive with regular SL analyses, the potential for expanding this to completely automated high-end SL analyses is evident. 
We hope the present work will constitute another step towards fully automatising SL analyses, as a standard tool for studying cluster mass distributions of very large samples, with the \emph{HST} and successors such as \emph{JWST}, and particularly, \emph{Euclid} and \emph{WFIRST}.

Finally we wish to highlight that the multiple-image automated finding process takes no more than a few minutes on a normal CPU. A second iteration that includes a refinement of the first lens model with the newly found multiple images, can take up to several hours due to the multi-dimensional minimisation.

\section*{acknowledgements}
The authors are grateful for useful discussions with Matthias Bartelmann. This work was supported in part by the transregional collaborative research centre TR 33 ``The Dark Universe" of the German Science Foundation.

\bibliographystyle{mnras}
\bibliography{outDan2Agnese}

\begin{thebibliography}{}
\makeatletter
\relax
\def\mn@urlcharsother{\let\do\@makeother \do\$\do\&\do\#\do\^\do\_\do\%\do\~}
\def\mn@doi{\begingroup\mn@urlcharsother \@ifnextchar [ {\mn@doi@}
  {\mn@doi@[]}}
\def\mn@doi@[#1]#2{\def\@tempa{#1}\ifx\@tempa\@empty \href
  {http://dx.doi.org/#2} {doi:#2}\else \href {http://dx.doi.org/#2} {#1}\fi
  \endgroup}
\def\mn@eprint#1#2{\mn@eprint@#1:#2::\@nil}
\def\mn@eprint@arXiv#1{\href {http://arxiv.org/abs/#1} {{\tt arXiv:#1}}}
\def\mn@eprint@dblp#1{\href {http://dblp.uni-trier.de/rec/bibtex/#1.xml}
  {dblp:#1}}
\def\mn@eprint@#1:#2:#3:#4\@nil{\def\@tempa {#1}\def\@tempb {#2}\def\@tempc
  {#3}\ifx \@tempc \@empty \let \@tempc \@tempb \let \@tempb \@tempa \fi \ifx
  \@tempb \@empty \def\@tempb {arXiv}\fi \@ifundefined
  {mn@eprint@\@tempb}{\@tempb:\@tempc}{\expandafter \expandafter \csname
  mn@eprint@\@tempb\endcsname \expandafter{\@tempc}}}

\bibitem[\protect\citeauthoryear{{Acebron} et~al.,}{{Acebron}
  et~al.}{2018}]{Acebron2018}
{Acebron} A.,  et~al., 2018, \mn@doi [\apj] {10.3847/1538-4357/aabe29}, \href
  {https://ui.adsabs.harvard.edu/abs/2018ApJ...858...42A} {858, 42}

\bibitem[\protect\citeauthoryear{{Amendola}, {Kunz}  \& {Sapone}}{{Amendola}
  et~al.}{2008}]{Amendola2008}
{Amendola} L.,  {Kunz} M.,   {Sapone} D.,  2008, \mn@doi [\jcap]
  {10.1088/1475-7516/2008/04/013}, \href
  {http://adsabs.harvard.edu/abs/2008JCAP...04..013A} {4, 13}

\bibitem[\protect\citeauthoryear{{Bartelmann}}{{Bartelmann}}{2010}]{Bartelmann2010reviewB}
{Bartelmann} M.,  2010, \mn@doi [Classical and Quantum Gravity]
  {10.1088/0264-9381/27/23/233001}, \href
  {http://adsabs.harvard.edu/abs/2010CQGra..27w3001B} {27, 233001}

\bibitem[\protect\citeauthoryear{{Bekenstein}}{{Bekenstein}}{2012}]{Bekenstein2012ReviewTeVes}
{Bekenstein} J.~D.,  2012, preprint, \href
  {http://adsabs.harvard.edu/abs/2012arXiv1201.2759B} {} (\mn@eprint {arXiv}
  {1201.2759})

\bibitem[\protect\citeauthoryear{{Ben{\'{\i}}tez}}{{Ben{\'{\i}}tez}}{2000}]{Benitez2000}
{Ben{\'{\i}}tez} N.,  2000, \mn@doi [\apj] {10.1086/308947}, \href
  {http://adsabs.harvard.edu/abs/2000ApJ...536..571B} {536, 571}

\bibitem[\protect\citeauthoryear{{Ben{\'{\i}}tez} et~al.,}{{Ben{\'{\i}}tez}
  et~al.}{2004}]{Benitez2004}
{Ben{\'{\i}}tez} N.,  et~al., 2004, \mn@doi [\apjs] {10.1086/380120}, \href
  {http://adsabs.harvard.edu/abs/2004ApJS..150....1B} {150, 1}

\bibitem[\protect\citeauthoryear{{Bertin} \& {Arnouts}}{{Bertin} \&
  {Arnouts}}{1996}]{BertinArnouts1996Sextractor}
{Bertin} E.,  {Arnouts} S.,  1996, \aaps, \href
  {http://adsabs.harvard.edu/abs/1996A%26AS..117..393B} {117, 393}

\bibitem[\protect\citeauthoryear{{Bouwens} et~al.,}{{Bouwens}
  et~al.}{2009}]{Bouwens2009behindclusters}
{Bouwens} R.~J.,  et~al., 2009, \mn@doi [\apj] {10.1088/0004-637X/690/2/1764},
  \href {http://adsabs.harvard.edu/abs/2009ApJ...690.1764B} {690, 1764}

\bibitem[\protect\citeauthoryear{{Brada{\v c}} et~al.,}{{Brada{\v c}}
  et~al.}{2008}]{Bradac2008rxj1347}
{Brada{\v c}} M.,  et~al., 2008, \mn@doi [\apj] {10.1086/588377}, \href
  {http://adsabs.harvard.edu/abs/2008ApJ...681..187B} {681, 187}

\bibitem[\protect\citeauthoryear{{Brada{\v c}} et~al.,}{{Brada{\v c}}
  et~al.}{2012}]{Bradac2012highz}
{Brada{\v c}} M.,  et~al., 2012, \mn@doi [\apjl] {10.1088/2041-8205/755/1/L7},
  \href {http://adsabs.harvard.edu/abs/2012ApJ...755L...7B} {755, L7}

\bibitem[\protect\citeauthoryear{{Bradley} et~al.,}{{Bradley}
  et~al.}{2012}]{Bradley2011}
{Bradley} L.~D.,  et~al., 2012, \mn@doi [\apj] {10.1088/0004-637X/747/1/3},
  \href {http://adsabs.harvard.edu/abs/2012ApJ...747....3B} {747, 3}

\bibitem[\protect\citeauthoryear{{Broadhurst} \& {Barkana}}{{Broadhurst} \&
  {Barkana}}{2008}]{BroadhurstBarkana2008}
{Broadhurst} T.~J.,  {Barkana} R.,  2008, \mn@doi [\mnras]
  {10.1111/j.1365-2966.2008.13852.x}, \href
  {http://adsabs.harvard.edu/abs/2008MNRAS.390.1647B} {390, 1647}

\bibitem[\protect\citeauthoryear{{Broadhurst} et~al.,}{{Broadhurst}
  et~al.}{2005}]{Broadhurst2005a}
{Broadhurst} T.,  et~al., 2005, \mn@doi [\apj] {10.1086/426494}, \href
  {http://adsabs.harvard.edu/abs/2005ApJ...621...53B} {621, 53}

\bibitem[\protect\citeauthoryear{{Broadhurst}, {Umetsu}, {Medezinski}, {Oguri}
  \& {Rephaeli}}{{Broadhurst} et~al.}{2008}]{Broadhurst2008}
{Broadhurst} T.,  {Umetsu} K.,  {Medezinski} E.,  {Oguri} M.,   {Rephaeli} Y.,
  2008, \mn@doi [\apjl] {10.1086/592400}, \href
  {http://adsabs.harvard.edu/abs/2008ApJ...685L...9B} {685, L9}

\bibitem[\protect\citeauthoryear{{Caminha} et~al.,}{{Caminha}
  et~al.}{2017}]{Caminha2017M0416}
{Caminha} G.~B.,  et~al., 2017, \mn@doi [\aap] {10.1051/0004-6361/201629297},
  \href {https://ui.adsabs.harvard.edu/abs/2017A&A...600A..90C} {600, A90}

\bibitem[\protect\citeauthoryear{{Caminha} et~al.,}{{Caminha}
  et~al.}{2019}]{Caminha2019_clash}
{Caminha} G.~B.,  et~al., 2019, arXiv e-prints, \href
  {https://ui.adsabs.harvard.edu/abs/2019arXiv190305103C} {p. arXiv:1903.05103}

\bibitem[\protect\citeauthoryear{{Cerny} et~al.,}{{Cerny}
  et~al.}{2018}]{Cerny2018}
{Cerny} C.,  et~al., 2018, \mn@doi [\apj] {10.3847/1538-4357/aabe7b}, \href
  {https://ui.adsabs.harvard.edu/abs/2018ApJ...859..159C} {859, 159}

\bibitem[\protect\citeauthoryear{{Chongchitnan} \& {Silk}}{{Chongchitnan} \&
  {Silk}}{2012}]{ChongchitnanSilk2012}
{Chongchitnan} S.,  {Silk} J.,  2012, \mn@doi [\prd]
  {10.1103/PhysRevD.85.063508}, \href
  {http://adsabs.harvard.edu/abs/2012PhRvD..85f3508C} {85, 063508}

\bibitem[\protect\citeauthoryear{{Coe}, {Ben{\'{\i}}tez}, {S{\'a}nchez}, {Jee},
  {Bouwens}  \& {Ford}}{{Coe} et~al.}{2006}]{Coe2006}
{Coe} D.,  {Ben{\'{\i}}tez} N.,  {S{\'a}nchez} S.~F.,  {Jee} M.,  {Bouwens} R.,
    {Ford} H.,  2006, \mn@doi [\aj] {10.1086/505530}, \href
  {http://adsabs.harvard.edu/abs/2006AJ....132..926C} {132, 926}

\bibitem[\protect\citeauthoryear{{Coe}, {Fuselier}, {Ben{\'{\i}}tez},
  {Broadhurst}, {Frye}  \& {Ford}}{{Coe} et~al.}{2008}]{Coe2008LensPerfect}
{Coe} D.,  {Fuselier} E.,  {Ben{\'{\i}}tez} N.,  {Broadhurst} T.,  {Frye} B.,
  {Ford} H.,  2008, \mn@doi [\apj] {10.1086/588250}, \href
  {http://adsabs.harvard.edu/abs/2008ApJ...681..814C} {681, 814}

\bibitem[\protect\citeauthoryear{{Coe}, {Ben{\'{\i}}tez}, {Broadhurst}  \&
  {Moustakas}}{{Coe} et~al.}{2010}]{Coe2010}
{Coe} D.,  {Ben{\'{\i}}tez} N.,  {Broadhurst} T.,   {Moustakas} L.~A.,  2010,
  \mn@doi [\apj] {10.1088/0004-637X/723/2/1678}, \href
  {http://adsabs.harvard.edu/abs/2010ApJ...723.1678C} {723, 1678}

\bibitem[\protect\citeauthoryear{{Coe} et~al.,}{{Coe}
  et~al.}{2013}]{Coe2012highz}
{Coe} D.,  et~al., 2013, \mn@doi [\apj] {10.1088/0004-637X/762/1/32}, \href
  {http://adsabs.harvard.edu/abs/2013ApJ...762...32C} {762, 32}

\bibitem[\protect\citeauthoryear{{Coe} et~al.,}{{Coe}
  et~al.}{2019}]{Coe2019arXivRELICS}
{Coe} D.,  et~al., 2019, preprint, \href
  {https://ui.adsabs.harvard.edu/abs/2019arXiv190302002C} {arXiv:1903.02002,
  arXiv} (\mn@eprint {arXiv} {1903.02002})

\bibitem[\protect\citeauthoryear{{Dalal}, {Holder}  \& {Hennawi}}{{Dalal}
  et~al.}{2004}]{Dalal+2004arcs}
{Dalal} N.,  {Holder} G.,   {Hennawi} J.~F.,  2004, \mn@doi [\apj]
  {10.1086/420960}, \href {http://adsabs.harvard.edu/abs/2004ApJ...609...50D}
  {609, 50}

\bibitem[\protect\citeauthoryear{{Diego}, {Protopapas}, {Sandvik}  \&
  {Tegmark}}{{Diego} et~al.}{2005}]{Diego2005Nonparam}
{Diego} J.~M.,  {Protopapas} P.,  {Sandvik} H.~B.,   {Tegmark} M.,  2005,
  \mn@doi [\mnras] {10.1111/j.1365-2966.2005.09021.x}, \href
  {http://adsabs.harvard.edu/abs/2005MNRAS.360..477D} {360, 477}

\bibitem[\protect\citeauthoryear{{Diego}, {Broadhurst}, {Zitrin}, {Lam}, {Lim},
  {Ford}  \& {Zheng}}{{Diego} et~al.}{2015}]{Diego2014M0717}
{Diego} J.~M.,  {Broadhurst} T.,  {Zitrin} A.,  {Lam} D.,  {Lim} J.,  {Ford}
  H.~C.,   {Zheng} W.,  2015, \mn@doi [\mnras] {10.1093/mnras/stv1168}, \href
  {http://adsabs.harvard.edu/abs/2015MNRAS.451.3920D} {451, 3920}

\bibitem[\protect\citeauthoryear{{Diego} et~al.,}{{Diego}
  et~al.}{2016}]{Diego2016refsdal}
{Diego} J.~M.,  et~al., 2016, \mn@doi [\mnras] {10.1093/mnras/stv2638}, \href
  {http://adsabs.harvard.edu/abs/2016MNRAS.456..356D} {456, 356}

\bibitem[\protect\citeauthoryear{{Ebeling}, {Edge}, {Mantz}, {Barrett},
  {Henry}, {Ma}  \& {van Speybroeck}}{{Ebeling}
  et~al.}{2010}]{Ebeling2010FinalMACS}
{Ebeling} H.,  {Edge} A.~C.,  {Mantz} A.,  {Barrett} E.,  {Henry} J.~P.,  {Ma}
  C.~J.,   {van Speybroeck} L.,  2010, \mn@doi [\mnras]
  {10.1111/j.1365-2966.2010.16920.x}, \href
  {http://adsabs.harvard.edu/abs/2010MNRAS.407...83E} {407, 83}

\bibitem[\protect\citeauthoryear{{Einasto}}{{Einasto}}{2009}]{Einasto2009DMreview}
{Einasto} J.,  2009, preprint, \href
  {http://esoads.eso.org/abs/2009arXiv0901.0632E} {0901.0632} (\mn@eprint
  {arXiv} {0901.0632})

\bibitem[\protect\citeauthoryear{{Emami}, {Broadhurst}, {Jimeno}, {Smoot},
  {Angulo}, {Lim}, {Chung Chu}  \& {Lazkoz}}{{Emami}
  et~al.}{2017}]{Emami2017NeutrinosClusters}
{Emami} R.,  {Broadhurst} T.,  {Jimeno} P.,  {Smoot} G.,  {Angulo} R.,  {Lim}
  J.,  {Chung Chu} M.,   {Lazkoz} R.,  2017, arXiv, \href
  {http://adsabs.harvard.edu/abs/2017arXiv171105210E} {1711.05210}

\bibitem[\protect\citeauthoryear{{Fedeli} \& {Bartelmann}}{{Fedeli} \&
  {Bartelmann}}{2007}]{Fedeli2007EDE1}
{Fedeli} C.,  {Bartelmann} M.,  2007, \mn@doi [\aap]
  {10.1051/0004-6361:20065976}, \href
  {http://adsabs.harvard.edu/abs/2007A%26A...461...49F} {461, 49}

\bibitem[\protect\citeauthoryear{{Franx}, {Illingworth}, {Kelson}, {van Dokkum}
   \& {Tran}}{{Franx} et~al.}{1997}]{Franx1997}
{Franx} M.,  {Illingworth} G.~D.,  {Kelson} D.~D.,  {van Dokkum} P.~G.,
  {Tran} K.-V.,  1997, \mn@doi [\apjl] {10.1086/310844}, \href
  {http://adsabs.harvard.edu/abs/1997ApJ...486L..75F} {486, L75}

\bibitem[\protect\citeauthoryear{{Gralla} et~al.,}{{Gralla}
  et~al.}{2011}]{Gralla2011}
{Gralla} M.~B.,  et~al., 2011, \mn@doi [\apj] {10.1088/0004-637X/737/2/74},
  \href {http://adsabs.harvard.edu/abs/2011ApJ...737...74G} {737, 74}

\bibitem[\protect\citeauthoryear{{Grillo} et~al.,}{{Grillo}
  et~al.}{2015}]{Grillo2015_0416}
{Grillo} C.,  et~al., 2015, \mn@doi [\apj] {10.1088/0004-637X/800/1/38}, \href
  {http://adsabs.harvard.edu/abs/2015ApJ...800...38G} {800, 38}

\bibitem[\protect\citeauthoryear{{Halkola}, {Seitz}  \& {Pannella}}{{Halkola}
  et~al.}{2006}]{Halkola2006}
{Halkola} A.,  {Seitz} S.,   {Pannella} M.,  2006, \mn@doi [\mnras]
  {10.1111/j.1365-2966.2006.10948.x}, \href
  {http://adsabs.harvard.edu/abs/2006MNRAS.372.1425H} {372, 1425}

\bibitem[\protect\citeauthoryear{{Hashimoto} et~al.,}{{Hashimoto}
  et~al.}{2018}]{Hashimoto2018}
{Hashimoto} T.,  et~al., 2018, \mn@doi [\nat] {10.1038/s41586-018-0117-z},
  \href {http://adsabs.harvard.edu/abs/2018Natur.557..392H} {557, 392}

\bibitem[\protect\citeauthoryear{{Hennawi}, {Dalal}, {Bode}  \&
  {Ostriker}}{{Hennawi} et~al.}{2007}]{Hennawi2007}
{Hennawi} J.~F.,  {Dalal} N.,  {Bode} P.,   {Ostriker} J.~P.,  2007, \mn@doi
  [\apj] {10.1086/497362}, \href
  {http://adsabs.harvard.edu/abs/2007ApJ...654..714H} {654, 714}

\bibitem[\protect\citeauthoryear{{Hoekstra}, {Herbonnet}, {Muzzin}, {Babul},
  {Mahdavi}, {Viola}  \& {Cacciato}}{{Hoekstra}
  et~al.}{2015}]{Hoekstra2015CCCP}
{Hoekstra} H.,  {Herbonnet} R.,  {Muzzin} A.,  {Babul} A.,  {Mahdavi} A.,
  {Viola} M.,   {Cacciato} M.,  2015, \mn@doi [\mnras] {10.1093/mnras/stv275},
  \href {https://ui.adsabs.harvard.edu/abs/2015MNRAS.449..685H} {449, 685}

\bibitem[\protect\citeauthoryear{{Host}}{{Host}}{2012}]{Host2012LOS}
{Host} O.,  2012, \mn@doi [\mnras] {10.1111/j.1745-3933.2011.01184.x}, \href
  {http://adsabs.harvard.edu/abs/2012MNRAS.420L..18H} {420, L18}

\bibitem[\protect\citeauthoryear{{Ishigaki}, {Kawamata}, {Ouchi}, {Oguri},
  {Shimasaku}  \& {Ono}}{{Ishigaki} et~al.}{2015}]{Ishigaki2014}
{Ishigaki} M.,  {Kawamata} R.,  {Ouchi} M.,  {Oguri} M.,  {Shimasaku} K.,
  {Ono} Y.,  2015, \mn@doi [\apj] {10.1088/0004-637X/799/1/12}, \href
  {http://adsabs.harvard.edu/abs/2015ApJ...799...12I} {799, 12}

\bibitem[\protect\citeauthoryear{{Jauzac} et~al.,}{{Jauzac}
  et~al.}{2015}]{Jauzac2015A2744}
{Jauzac} M.,  et~al., 2015, \mn@doi [\mnras] {10.1093/mnras/stv1402}, \href
  {http://adsabs.harvard.edu/abs/2015MNRAS.452.1437J} {452, 1437}

\bibitem[\protect\citeauthoryear{{Johnson}, {Sharon}, {Bayliss}, {Gladders},
  {Coe}  \& {Ebeling}}{{Johnson} et~al.}{2014}]{Johnson2014HFFmodels}
{Johnson} T.~L.,  {Sharon} K.,  {Bayliss} M.~B.,  {Gladders} M.~D.,  {Coe} D.,
   {Ebeling} H.,  2014, \mn@doi [\apj] {10.1088/0004-637X/797/1/48}, \href
  {http://adsabs.harvard.edu/abs/2014ApJ...797...48J} {797, 48}

\bibitem[\protect\citeauthoryear{{Jullo}, {Kneib}, {Limousin},
  {El{\'{\i}}asd{\'o}ttir}, {Marshall}  \& {Verdugo}}{{Jullo}
  et~al.}{2007}]{Jullo2007Lenstool}
{Jullo} E.,  {Kneib} J.-P.,  {Limousin} M.,  {El{\'{\i}}asd{\'o}ttir} {\'A}.,
  {Marshall} P.~J.,   {Verdugo} T.,  2007, \mn@doi [New Journal of Physics]
  {10.1088/1367-2630/9/12/447}, \href
  {http://adsabs.harvard.edu/abs/2007NJPh....9..447J} {9, 447}

\bibitem[\protect\citeauthoryear{{Jullo}, {Natarajan}, {Kneib}, {D'Aloisio},
  {Limousin}, {Richard}  \& {Schimd}}{{Jullo} et~al.}{2010}]{Jullo2010}
{Jullo} E.,  {Natarajan} P.,  {Kneib} J.,  {D'Aloisio} A.,  {Limousin} M.,
  {Richard} J.,   {Schimd} C.,  2010, \mn@doi [Science]
  {10.1126/science.1185759}, \href
  {http://adsabs.harvard.edu/abs/2010Sci...329..924J} {329, 924}

\bibitem[\protect\citeauthoryear{{Kawamata}, {Oguri}, {Ishigaki}, {Shimasaku}
  \& {Ouchi}}{{Kawamata} et~al.}{2016}]{Kawamata2016modelsHFF}
{Kawamata} R.,  {Oguri} M.,  {Ishigaki} M.,  {Shimasaku} K.,   {Ouchi} M.,
  2016, \mn@doi [\apj] {10.3847/0004-637X/819/2/114}, \href
  {http://adsabs.harvard.edu/abs/2016ApJ...819..114K} {819, 114}

\bibitem[\protect\citeauthoryear{{Keeton}}{{Keeton}}{2001}]{Keeton2001models}
{Keeton} C.~R.,  2001, ArXiv Astrophysics e-prints, \href
  {http://adsabs.harvard.edu/abs/2001astro.ph..2340K} {0102340}

\bibitem[\protect\citeauthoryear{{Kneib} \& {Natarajan}}{{Kneib} \&
  {Natarajan}}{2011}]{Kneib2011review}
{Kneib} J.-P.,  {Natarajan} P.,  2011, \mn@doi [\aapr]
  {10.1007/s00159-011-0047-3}, \href
  {http://adsabs.harvard.edu/abs/2011A%26ARv..19...47K} {19, 47}

\bibitem[\protect\citeauthoryear{{Kneib}, {Mellier}, {Fort}  \&
  {Mathez}}{{Kneib} et~al.}{1993}]{Kneib1993Lenstool}
{Kneib} J.~P.,  {Mellier} Y.,  {Fort} B.,   {Mathez} G.,  1993, \aap, \href
  {http://adsabs.harvard.edu/abs/1993A%26A...273..367K} {273, 367}

\bibitem[\protect\citeauthoryear{{Kneib}, {Ellis}, {Santos}  \&
  {Richard}}{{Kneib} et~al.}{2004}]{Kneib2004z7}
{Kneib} J.-P.,  {Ellis} R.~S.,  {Santos} M.~R.,   {Richard} J.,  2004, \mn@doi
  [\apj] {10.1086/386281}, \href
  {http://adsabs.harvard.edu/abs/2004ApJ...607..697K} {607, 697}

\bibitem[\protect\citeauthoryear{{Koekemoer} et~al.,}{{Koekemoer}
  et~al.}{2002}]{Koekemoer2002}
{Koekemoer} A.~M.,  et~al., 2002, \mn@doi [\apj] {10.1086/338129}, \href
  {http://adsabs.harvard.edu/abs/2002ApJ...567..657K} {567, 657}

\bibitem[\protect\citeauthoryear{{Koekemoer} et~al.,}{{Koekemoer}
  et~al.}{2011}]{Koekemoer2011}
{Koekemoer} A.~M.,  et~al., 2011, \mn@doi [\apjs] {10.1088/0067-0049/197/2/36},
  \href {http://adsabs.harvard.edu/abs/2011ApJS..197...36K} {197, 36}

\bibitem[\protect\citeauthoryear{{Komatsu} et~al.,}{{Komatsu}
  et~al.}{2011}]{Komatsu2011WMAP7}
{Komatsu} E.,  et~al., 2011, \mn@doi [\apjs] {10.1088/0067-0049/192/2/18},
  \href {http://esoads.eso.org/abs/2011ApJS..192...18K} {192, 18}

\bibitem[\protect\citeauthoryear{{Kroupa}}{{Kroupa}}{2012}]{Kroupa2012problemDM}
{Kroupa} P.,  2012, \mn@doi [\pasa] {10.1071/AS12005}, \href
  {http://adsabs.harvard.edu/abs/2012PASA...29..395K} {29, 395}

\bibitem[\protect\citeauthoryear{{Liesenborgs}, {De Rijcke}  \&
  {Dejonghe}}{{Liesenborgs} et~al.}{2006}]{Liesenborgs2006}
{Liesenborgs} J.,  {De Rijcke} S.,   {Dejonghe} H.,  2006, \mn@doi [\mnras]
  {10.1111/j.1365-2966.2006.10040.x}, \href
  {http://adsabs.harvard.edu/abs/2006MNRAS.367.1209L} {367, 1209}

\bibitem[\protect\citeauthoryear{{Liesenborgs}, {de Rijcke}, {Dejonghe}  \&
  {Bekaert}}{{Liesenborgs} et~al.}{2007}]{Liesenborgs2007}
{Liesenborgs} J.,  {de Rijcke} S.,  {Dejonghe} H.,   {Bekaert} P.,  2007,
  \mn@doi [\mnras] {10.1111/j.1365-2966.2007.12236.x}, \href
  {http://adsabs.harvard.edu/abs/2007MNRAS.380.1729L} {380, 1729}

\bibitem[\protect\citeauthoryear{{Liesenborgs}, {de Rijcke}, {Dejonghe}  \&
  {Bekaert}}{{Liesenborgs} et~al.}{2008}]{Liesenborgs2008CL0024L}
{Liesenborgs} J.,  {de Rijcke} S.,  {Dejonghe} H.,   {Bekaert} P.,  2008,
  \mn@doi [\mnras] {10.1111/j.1365-2966.2008.13586.x}, \href
  {http://adsabs.harvard.edu/abs/2008MNRAS.389..415L} {389, 415}

\bibitem[\protect\citeauthoryear{{Lotz} et~al.,}{{Lotz}
  et~al.}{2017}]{Lotz2016HFF}
{Lotz} J.~M.,  et~al., 2017, \mn@doi [\apj] {10.3847/1538-4357/837/1/97}, \href
  {http://adsabs.harvard.edu/abs/2017ApJ...837...97L} {837, 97}

\bibitem[\protect\citeauthoryear{{Meneghetti}, {Rasia}, {Merten}, {Bellagamba},
  {Ettori}, {Mazzotta}, {Dolag}  \& {Marri}}{{Meneghetti}
  et~al.}{2010}]{Meneghetti2010b}
{Meneghetti} M.,  {Rasia} E.,  {Merten} J.,  {Bellagamba} F.,  {Ettori} S.,
  {Mazzotta} P.,  {Dolag} K.,   {Marri} S.,  2010, \mn@doi [\aap]
  {10.1051/0004-6361/200913222}, \href
  {http://adsabs.harvard.edu/abs/2010A%26A...514A..93M} {514, A93+}

\bibitem[\protect\citeauthoryear{{Meneghetti} et~al.,}{{Meneghetti}
  et~al.}{2017}]{Meneghetti2016SIMSCOMP}
{Meneghetti} M.,  et~al., 2017, \mn@doi [\mnras] {10.1093/mnras/stx2064}, \href
  {http://adsabs.harvard.edu/abs/2017MNRAS.472.3177M} {472, 3177}

\bibitem[\protect\citeauthoryear{{Merten} et~al.,}{{Merten}
  et~al.}{2015}]{Merten2014CLASHcM}
{Merten} J.,  et~al., 2015, \mn@doi [\apj] {10.1088/0004-637X/806/1/4}, \href
  {http://adsabs.harvard.edu/abs/2015ApJ...806....4M} {806, 4}

\bibitem[\protect\citeauthoryear{{Milgrom}}{{Milgrom}}{1983}]{Milgrom1983MOND}
{Milgrom} M.,  1983, \mn@doi [\apj] {10.1086/161130}, \href
  {http://adsabs.harvard.edu/abs/1983ApJ...270..365M} {270, 365}

\bibitem[\protect\citeauthoryear{{Molino} et~al.,}{{Molino}
  et~al.}{2017}]{Molino2017CLASHcats}
{Molino} A.,  et~al., 2017, \mn@doi [\mnras] {10.1093/mnras/stx1243}, \href
  {https://ui.adsabs.harvard.edu/abs/2017MNRAS.470...95M} {470, 95}

\bibitem[\protect\citeauthoryear{{Navarro}, {Frenk}  \& {White}}{{Navarro}
  et~al.}{1996}]{Navarro1996}
{Navarro} J.~F.,  {Frenk} C.~S.,   {White} S.~D.~M.,  1996, \mn@doi [\apj]
  {10.1086/177173}, \href {http://adsabs.harvard.edu/abs/1996ApJ...462..563N}
  {462, 563}

\bibitem[\protect\citeauthoryear{{Newman}, {Treu}, {Ellis}, {Sand}, {Nipoti},
  {Richard}  \& {Jullo}}{{Newman} et~al.}{2013}]{Newman2013}
{Newman} A.~B.,  {Treu} T.,  {Ellis} R.~S.,  {Sand} D.~J.,  {Nipoti} C.,
  {Richard} J.,   {Jullo} E.,  2013, \mn@doi [\apj]
  {10.1088/0004-637X/765/1/24}, \href
  {http://adsabs.harvard.edu/abs/2013ApJ...765...24N} {765, 24}

\bibitem[\protect\citeauthoryear{{Oguri} \& {Blandford}}{{Oguri} \&
  {Blandford}}{2009}]{OguriBlandford2009}
{Oguri} M.,  {Blandford} R.~D.,  2009, \mn@doi [\mnras]
  {10.1111/j.1365-2966.2008.14154.x}, \href
  {http://adsabs.harvard.edu/abs/2009MNRAS.392..930O} {392, 930}

\bibitem[\protect\citeauthoryear{{Oguri}, {Bayliss}, {Dahle}, {Sharon},
  {Gladders}, {Natarajan}, {Hennawi}  \& {Koester}}{{Oguri}
  et~al.}{2012}]{Oguri201238clusters}
{Oguri} M.,  {Bayliss} M.~B.,  {Dahle} H.,  {Sharon} K.,  {Gladders} M.~D.,
  {Natarajan} P.,  {Hennawi} J.~F.,   {Koester} B.~P.,  2012, \mn@doi [\mnras]
  {10.1111/j.1365-2966.2011.20248.x}, \href
  {http://adsabs.harvard.edu/abs/2012MNRAS.420.3213O} {420, 3213}

\bibitem[\protect\citeauthoryear{{Oguri} et~al.,}{{Oguri}
  et~al.}{2013}]{Oguri2012SL}
{Oguri} M.,  et~al., 2013, \mn@doi [\mnras] {10.1093/mnras/sts351}, \href
  {http://adsabs.harvard.edu/abs/2013MNRAS.429..482O} {429, 482}

\bibitem[\protect\citeauthoryear{{Okabe} \& {Smith}}{{Okabe} \&
  {Smith}}{2016}]{Okabe2016cM}
{Okabe} N.,  {Smith} G.~P.,  2016, \mn@doi [\mnras] {10.1093/mnras/stw1539},
  \href {https://ui.adsabs.harvard.edu/abs/2016MNRAS.461.3794O} {461, 3794}

\bibitem[\protect\citeauthoryear{{Planck Collaboration} et~al.,}{{Planck
  Collaboration} et~al.}{2015}]{Planck2015catalog}
{Planck Collaboration} et~al., 2015, \mn@doi [\aap]
  {10.1051/0004-6361/201525787}, \href
  {http://adsabs.harvard.edu/abs/2015A%26A...581A..14P} {581, A14}

\bibitem[\protect\citeauthoryear{{Planck Collaboration} et~al.,}{{Planck
  Collaboration} et~al.}{2018}]{Planck2018Params}
{Planck Collaboration} et~al., 2018, arXiv, \href
  {http://adsabs.harvard.edu/abs/2018arXiv180706209P} {1807.06209}

\bibitem[\protect\citeauthoryear{{Ponente} \& {Diego}}{{Ponente} \&
  {Diego}}{2011}]{PonenteDiego2011}
{Ponente} P.~P.,  {Diego} J.~M.,  2011, \mn@doi [\aap]
  {10.1051/0004-6361/201117382}, \href
  {http://adsabs.harvard.edu/abs/2011A%26A...535A.119P} {535, A119}

\bibitem[\protect\citeauthoryear{{Postman} et~al.,}{{Postman}
  et~al.}{2012}]{PostmanCLASHoverview}
{Postman} M.,  et~al., 2012, \mn@doi [\apjs] {10.1088/0067-0049/199/2/25},
  \href {http://adsabs.harvard.edu/abs/2012ApJS..199...25P} {199, 25}

\bibitem[\protect\citeauthoryear{{Puchwein} \& {Hilbert}}{{Puchwein} \&
  {Hilbert}}{2009}]{PuchweinHilbert2009}
{Puchwein} E.,  {Hilbert} S.,  2009, \mn@doi [\mnras]
  {10.1111/j.1365-2966.2009.15227.x}, \href
  {http://adsabs.harvard.edu/abs/2009MNRAS.398.1298P} {398, 1298}

\bibitem[\protect\citeauthoryear{{Redlich}, {Bolejko}, {Meyer}, {Lewis}  \&
  {Bartelmann}}{{Redlich} et~al.}{2014}]{Redlich2014LTB}
{Redlich} M.,  {Bolejko} K.,  {Meyer} S.,  {Lewis} G.~F.,   {Bartelmann} M.,
  2014, \mn@doi [\aap] {10.1051/0004-6361/201424553}, \href
  {http://adsabs.harvard.edu/abs/2014A%26A...570A..63R} {570, A63}

\bibitem[\protect\citeauthoryear{{Richard} et~al.,}{{Richard}
  et~al.}{2010}]{Richard2010locuss20}
{Richard} J.,  et~al., 2010, \mn@doi [\mnras]
  {10.1111/j.1365-2966.2009.16274.x}, \href
  {http://adsabs.harvard.edu/abs/2010MNRAS.404..325R} {404, 325}

\bibitem[\protect\citeauthoryear{{Richard}, {Kneib}, {Ebeling}, {Stark},
  {Egami}  \& {Fiedler}}{{Richard} et~al.}{2011}]{Richard2011}
{Richard} J.,  {Kneib} J.-P.,  {Ebeling} H.,  {Stark} D.~P.,  {Egami} E.,
  {Fiedler} A.~K.,  2011, \mn@doi [\mnras] {10.1111/j.1745-3933.2011.01050.x},
  \href {http://adsabs.harvard.edu/abs/2011MNRAS.414L..31R} {414, L31}

\bibitem[\protect\citeauthoryear{{Sadeh} \& {Rephaeli}}{{Sadeh} \&
  {Rephaeli}}{2008}]{SadehRephaeli2008}
{Sadeh} S.,  {Rephaeli} Y.,  2008, \mn@doi [\mnras]
  {10.1111/j.1365-2966.2008.13501.x}, \href
  {http://adsabs.harvard.edu/abs/2008MNRAS.388.1759S} {388, 1759}

\bibitem[\protect\citeauthoryear{{Seidel} \& {Bartelmann}}{{Seidel} \&
  {Bartelmann}}{2007}]{Seidel2007Arcfinder}
{Seidel} G.,  {Bartelmann} M.,  2007, \mn@doi [\aap]
  {10.1051/0004-6361:20066097}, \href
  {http://adsabs.harvard.edu/abs/2007A%26A...472..341S} {472, 341}

\bibitem[\protect\citeauthoryear{{Sereno}, {Jetzer}  \& {Lubini}}{{Sereno}
  et~al.}{2010}]{Sereno2010}
{Sereno} M.,  {Jetzer} P.,   {Lubini} M.,  2010, \mn@doi [\mnras]
  {10.1111/j.1365-2966.2010.16248.x}, \href
  {http://adsabs.harvard.edu/abs/2010MNRAS.403.2077S} {403, 2077}

\bibitem[\protect\citeauthoryear{{Smith}, {Kneib}, {Smail}, {Mazzotta},
  {Ebeling}  \& {Czoske}}{{Smith} et~al.}{2005}]{Smith2005}
{Smith} G.~P.,  {Kneib} J.,  {Smail} I.,  {Mazzotta} P.,  {Ebeling} H.,
  {Czoske} O.,  2005, \mn@doi [\mnras] {10.1111/j.1365-2966.2005.08911.x},
  \href {http://adsabs.harvard.edu/abs/2005MNRAS.359..417S} {359, 417}

\bibitem[\protect\citeauthoryear{{Stapelberg}, {Carrasco}  \&
  {Maturi}}{{Stapelberg} et~al.}{2019}]{Stapelberg2019}
{Stapelberg} S.,  {Carrasco} M.,   {Maturi} M.,  2019, \mn@doi [\mnras]
  {10.1093/mnras/sty2784}, \href
  {https://ui.adsabs.harvard.edu/abs/2019MNRAS.482.1824S} {482, 1824}

\bibitem[\protect\citeauthoryear{{Tian}, {Ko}  \& {Chiu}}{{Tian}
  et~al.}{2013}]{Tian2012TeVesLensing}
{Tian} Y.,  {Ko} C.-M.,   {Chiu} M.-C.,  2013, \mn@doi [\apj]
  {10.1088/0004-637X/770/2/154}, \href
  {https://ui.adsabs.harvard.edu/abs/2013ApJ...770..154T} {770, 154}

\bibitem[\protect\citeauthoryear{{Treu} et~al.,}{{Treu}
  et~al.}{2016}]{Treu2016Refsdal}
{Treu} T.,  et~al., 2016, \mn@doi [\apj] {10.3847/0004-637X/817/1/60}, \href
  {http://adsabs.harvard.edu/abs/2016ApJ...817...60T} {817, 60}

\bibitem[\protect\citeauthoryear{{Umetsu}, {Broadhurst}, {Zitrin}, {Medezinski}
   \& {Hsu}}{{Umetsu} et~al.}{2011}]{Umetsu2011}
{Umetsu} K.,  {Broadhurst} T.,  {Zitrin} A.,  {Medezinski} E.,   {Hsu} L.,
  2011, \mn@doi [\apj] {10.1088/0004-637X/729/2/127}, \href
  {http://adsabs.harvard.edu/abs/2011ApJ...729..127U} {729, 127}

\bibitem[\protect\citeauthoryear{{Umetsu} et~al.,}{{Umetsu}
  et~al.}{2014}]{Umetsu2014CLASH_WL}
{Umetsu} K.,  et~al., 2014, \mn@doi [\apj] {10.1088/0004-637X/795/2/163}, \href
  {http://adsabs.harvard.edu/abs/2014ApJ...795..163U} {795, 163}

\bibitem[\protect\citeauthoryear{{Wambsganss}, {Cen}, {Ostriker}  \&
  {Turner}}{{Wambsganss} et~al.}{1995}]{Wambsganss1995Testcdm}
{Wambsganss} J.,  {Cen} R.,  {Ostriker} J.~P.,   {Turner} E.~L.,  1995, \mn@doi
  [Science] {10.1126/science.268.5208.274}, \href
  {http://adsabs.harvard.edu/abs/1995Sci...268..274W} {268, 274}

\bibitem[\protect\citeauthoryear{{Williams} \& {Saha}}{{Williams} \&
  {Saha}}{2011}]{WilliamsSaha2011Offset}
{Williams} L.~L.~R.,  {Saha} P.,  2011, \mn@doi [\mnras]
  {10.1111/j.1365-2966.2011.18716.x}, \href
  {http://adsabs.harvard.edu/abs/2011MNRAS.415..448W} {415, 448}

\bibitem[\protect\citeauthoryear{{Wong}, {Ammons}, {Keeton}  \&
  {Zabludoff}}{{Wong} et~al.}{2012}]{Wong2012OptLenses}
{Wong} K.~C.,  {Ammons} S.~M.,  {Keeton} C.~R.,   {Zabludoff} A.~I.,  2012,
  \mn@doi [\apj] {10.1088/0004-637X/752/2/104}, \href
  {http://adsabs.harvard.edu/abs/2012ApJ...752..104W} {752, 104}

\bibitem[\protect\citeauthoryear{{Zackrisson} et~al.,}{{Zackrisson}
  et~al.}{2012}]{Zackrisson2012}
{Zackrisson} E.,  et~al., 2012, \mn@doi [\mnras]
  {10.1111/j.1365-2966.2012.22078.x}, \href
  {http://adsabs.harvard.edu/abs/2012MNRAS.427.2212Z} {427, 2212}

\bibitem[\protect\citeauthoryear{{Zheng} et~al.,}{{Zheng}
  et~al.}{2009}]{Zheng2009}
{Zheng} W.,  et~al., 2009, \mn@doi [\apj] {10.1088/0004-637X/697/2/1907}, \href
  {http://adsabs.harvard.edu/abs/2009ApJ...697.1907Z} {697, 1907}

\bibitem[\protect\citeauthoryear{{Zitrin} \& {Broadhurst}}{{Zitrin} \&
  {Broadhurst}}{2009}]{ZitrinBroadhurst2009}
{Zitrin} A.,  {Broadhurst} T.,  2009, \mn@doi [\apjl]
  {10.1088/0004-637X/703/2/L132}, \href
  {http://adsabs.harvard.edu/abs/2009ApJ...703L.132Z} {703, L132}

\bibitem[\protect\citeauthoryear{{Zitrin} et~al.,}{{Zitrin}
  et~al.}{2009}]{Zitrin2009b}
{Zitrin} A.,  et~al., 2009, \mn@doi [\mnras]
  {10.1111/j.1365-2966.2009.14899.x}, \href
  {http://adsabs.harvard.edu/abs/2009MNRAS.396.1985Z} {396, 1985}

\bibitem[\protect\citeauthoryear{{Zitrin}, {Broadhurst}, {Bartelmann},
  {Rephaeli}, {Oguri}, {Ben{\'{\i}}tez}, {Hao}  \& {Umetsu}}{{Zitrin}
  et~al.}{2012a}]{Zitrin2011d}
{Zitrin} A.,  {Broadhurst} T.,  {Bartelmann} M.,  {Rephaeli} Y.,  {Oguri} M.,
  {Ben{\'{\i}}tez} N.,  {Hao} J.,   {Umetsu} K.,  2012a, \mn@doi [\mnras]
  {10.1111/j.1365-2966.2012.21041.x}, \href
  {http://adsabs.harvard.edu/abs/2012MNRAS.423.2308Z} {423, 2308}

\bibitem[\protect\citeauthoryear{{Zitrin} et~al.,}{{Zitrin}
  et~al.}{2012b}]{Zitrin2012CLASH0329}
{Zitrin} A.,  et~al., 2012b, \mn@doi [\apjl] {10.1088/2041-8205/747/1/L9},
  \href {http://adsabs.harvard.edu/abs/2012ApJ...747L...9Z} {747, L9}

\bibitem[\protect\citeauthoryear{{Zitrin} et~al.,}{{Zitrin}
  et~al.}{2015}]{Zitrin2015CLASH25}
{Zitrin} A.,  et~al., 2015, \mn@doi [\apj] {10.1088/0004-637X/801/1/44}, \href
  {http://adsabs.harvard.edu/abs/2015ApJ...801...44Z} {801, 44}

\bibitem[\protect\citeauthoryear{{von der Linden} et~al.,}{{von der Linden}
  et~al.}{2014}]{vonderLinden2014WTG}
{von der Linden} A.,  et~al., 2014, \mn@doi [\mnras] {10.1093/mnras/stt1945},
  \href {https://ui.adsabs.harvard.edu/abs/2014MNRAS.439....2V} {439, 2}

\makeatother
\end{thebibliography}

\appendix 

\section{The LTM model}\label{appendix:1}

We outline our revised version of the LTM formalism \citep{Broadhurst2005a, Zitrin2009b, Zitrin2011d} and its implementation for constructing an initial lens model. This is the same formalism recently used to analyse some RELICS \citep{Coe2019arXivRELICS} galaxy clusters \citep[e.g.][]{Acebron2018}. We assume that the mass distribution of each cluster member galaxy can be described by a surface-density
profile, $\Sigma(r)=\Sigma_{q0}r^{-q}$, with r being the radius from the centre of the galaxy and $\Sigma_{q0}$ some normalisation factor. This profile can be integrated to
give the interior mass, $M(<\theta)=\frac{2\pi
\Sigma_{q0}}{2-q}(d_{l}\theta)^{2-q}$, which results in a deflection angle, due to a single galaxy, of:

\begin{equation}
\label{deflectiona}
 \alpha(\theta)=\frac{4G\frac{2\pi
\Sigma_{q0}}{2-q}d_{l}^{~1-q}}{c^2}\frac{d_{ls}}{d_{s}}\theta^{1-q},
\end{equation}
where $\theta$ is the angular distance from the galaxy's centre, $G$ the gravitational constant, $c$ the speed of light, and $d_{l}$, $d_{s}$, $d_{ls}$ the angular diameter distances to the lens, to the source, and between the lens and the source. Simply put, $ \alpha(\theta)\propto\theta^{1-q}$, per given lens and source redshifts (or angular diameter distances).  Note that for all galaxies we assume circular symmetry, aside from one or two BCGs to which ellipticity can be incorporated. To these galaxies a core can also be added.

The deflection angle at a certain point $\vec{\theta}$
due to the lumpy galaxy components is
simply a linear superposition of all galaxy contributions scaled by
their luminosities, or fluxes $F_{i}$ (times some normalisation factor):
\begin{equation}
\label{deflection3}
 \vec{\alpha}_{gal}(\vec{\theta})\propto\sum_{i}
F_{i}\, |\vec{\theta}-\vec{\theta}_i|^{1-q}
\frac{\vec{\theta}-\vec{\theta}_i}{|\vec{\theta}-\vec{\theta}_i|}.
\end{equation}

In practice we use a discretised version of equation \ref{deflection3}, over a 2D square grid $\vec\theta_m$ of $N\times N$ pixels, given by:
\begin{equation}
\label{deflection_x1}
\alpha_{gal,x}(\vec\theta_m)=\sum_{i}
F_{i}\, \frac
{\Delta x_{mi}}{[(\Delta x_{mi})^2
 +
 (\Delta y_{mi})^2]^{q/2}},
 \end{equation}
\begin{equation}
\label{deflection_y1}
\alpha_{gal,y}(\vec\theta_m)=\sum_{i}
F_{i}\, \frac
{\Delta y_{mi}}{[(\Delta x_{mi})^2
 +
 (\Delta y_{mi})^2]^{q/2}},
 \end{equation}
where $(\Delta x_{mi},\Delta y_{mi})$ is the displacement vector
$\vec\theta_m-\vec\theta_i$ of the $m$th pixel point, with respect to the
$i$th galaxy position $\vec\theta_i$. Note that there should be some normalisation factor in the expression of the deflections fields, which we initially set to unity. 

In the code the sum is not performed explicitly as above, but instead is performed in Fourier space for speed-up purposes. The Fourier transform kernel is the same for all galaxies (substantially speeding up the calculation), and differs only for the independently modelled BCGs. The contribution of the BCGs can also be independently scaled. 

The mass density distribution from the galaxy component is then given by the divergence of this deflection field (times half). 

The second component of the model, the DM distribution, is assumed to be a Gaussian-smoothed version of the galaxy component. Therefore if $P$ represents the mass surface density (say, $\kappa$) of the smoothed map, then the deflection field from the DM component is simply:

\begin{equation}
\label{deflection_xDM}
\alpha_{DM,x}(\vec\theta_m)=\sum_{i}
P_i\, \frac
{\Delta x_{mi}}{[(\Delta x_{mi})^2
 +
 (\Delta y_{mi})^2]},
 \end{equation}
\begin{equation}
\label{deflection_yDM}
\alpha_{DM,y}(\vec\theta_m)=\sum_{i}
P_i\, \frac
{\Delta y_{mi}}{[(\Delta x_{mi})^2
 +
 (\Delta y_{mi})^2]},
 \end{equation}
where $P_{i}$ represents the (unnormalised) mass density value in the $i$th pixel of the smooth component as obtained by the smoothing procedure. 

Note that in practice, instead of smoothing the galaxy-component mass density distribution directly, the convolution with a Gaussian kernel and the relevant summation to a deflection field are performed in one quick step in Fourier space.

The third component of the model is an external shear, given by: 
\begin{equation}
\alpha_{ex,x}(\vec\theta_m)=\gamma\cos(2\phi)\Delta x_{m}+\gamma\sin(2\phi)\Delta y_{m}
\end{equation}
\begin{equation}
\alpha_{ex,y}(\vec\theta_m)=\gamma\sin(2\phi)\Delta x_{m}-\gamma\cos(2\phi)\Delta y_{m},
\end{equation}
where $\gamma$ is the strength of the external shear ($\gamma\geqslant0$), and $\phi$ its position angle.

The total deflection field is obtained by combining the components as follows:
\begin{equation}
\label{defTot}
\vec{\alpha}_T(\vec\theta)=K~[N_{1}K_{gal} \vec\alpha_{gal}(\vec\theta)+N_{2}(1-K_{gal})\vec\alpha_{DM}(\vec\theta)] + \vec\alpha_{ex}(\vec\theta),
\end{equation}
where $N_{1}$ and $N_{2}$ are normalisation factors placing the galaxy and DM deflection fields onto the same (arbitrary) scale, $K_{gal}$ is the relative contribution of the galaxy component to the deflection field, and K is the overall normalisation. Note that the above procedure is typically performed with a $4\times4$ lower resolution than the input HST images, and then the deflection fields are interpolated to the full HST resolution. 

A \textsc{MATLAB} implementation of the code is available online\footnote{\color{blue}{https://github.com/adizitrin/initialLTM}}. Typically, in our various analyses with this implementation we find that $q\sim1.3$, the Gaussian smoothing length is roughly $\sim100-200$ pixels, and $K_{gal}\sim0.05-0.1$. The external shear is about $\sim0.1$ on average for elliptical lenses, aligned typically along the main axis of the cluster -- which is usually also the BCG elongation direction. With these, it is easy to construct an initial lens model for any given cluster (see more details in \ref{pre_mass_model}). The only missing factor is the overall scaling. Since the overall normalisation is directly related to the size of the lens, or Einstein radius, a good initial guess can be such that a typical Einstein radius size is obtained (say $\sim15-20\arcsec$). In our implementation we use $N_{1}$ and $N_{2}$ such that the mean of the absolute value of each deflection field is scaled to an arbitrary value of 200 (pixels), and then $K$ varies between $K\sim0.5$ for the small lenses and $K\sim3$ for large lenses. In addition, a scaling between the galaxy luminosities and the normalisation factor $K$, as was done by \citep[][]{Zitrin2011d} for their older-version code, remains for future work (see also \citealt{Stapelberg2019}).

\section{Parameters used to run MIFAL}\label{appendix:2}
(see Tables on next page)
\begin{table*}

\caption{Parameters used to construct the preliminary lens models}\label{modelparams}

\centering
\resizebox{2.1\columnwidth}{!}{
\begin{tabular}{c c c c c c c c c c c c c c c c c c c c}
\toprule \toprule

Cluster & $q$ & $S$ &   $K$ &   $K_{gal}$ &  $\gamma_{ex}$ & $\phi_{ex}$ & BCG1 & BCG1 & BCG1 & BCG1  & BCG1  & BCG1  & BCG2 & BCG2 & BCG2 & BCG2 & BCG2  & BCG2   \\
 name & & (pixels) &    &  &   & $(^{\circ})$ & RA $(^{\circ})$& DEC $(^{\circ})$& weight & $r_{core}~[\arcsec]$&ellipticity & P.A. $(^{\circ})$ &  RA & DEC & weight  &   $r_{core}~[\arcsec]$&  ellipticity&  P.A. $(^{\circ})$ \\
\midrule
M0329 & 1.3 & 150 & 1.5 & 0.07 & 0 & 0 & 52.423219  & -2.1962308   & 5 & 0 & 0        & 0  &  52.413118 &  -2.1914246  & 3 & 0 & 0 & 0 \\
M1720 & 1.3 & 150 & 1.5 & 0.07 & 0 & 0 &  260.06980  &  35.607312  & 5 & 0 & 0.123 & 87.2 &   ---          & ---               &  --- &  --- &    ---  &   --- \\
M1931 & 1.3 & 150 & 1.2 & 0.07 & 0.1 & 82 & 292.95678 & -26.57572   & 5 & 0 & 0.5$^a$ & 82$^a$ &   --- & ---             &  --- &  --- &  ---& --- & \\
\bottomrule
\end{tabular}
}                                        

{\footnotesize\flushleft
Note - For explanation on the parameters see Appendix \ref{appendix:1}. \\
$Column~1:$ cluster name; $Column~2:$ galaxy power-law exponent; $Column~3:$ smoothing Gaussian width, in pixels, $Column~4:$ overall normalisation factor; $Column~5:$ contribution of galaxies relative to DM; $Column~6~\&~7:$ strength and direction (north of west) of external shear; $Column~8~\&~9:$ coordinates of the first BCG, J2000; $Column~10:$ weight of the second BCG, i.e., its mass-to-light ratio compared to normal cluster galaxies; $Columns~11,12~\&~13:$ core radius of the first BCG, its ellipticity and position angle (north of west); $Columns~14~\&~15:$ coordinates of the second BCG, J2000;  $Columns~15:$ weight of the second BCG, i.e., its mass-to-light ratio compared to normal cluster galaxies; $Columns~17,18~\&~19:$ core radius of the second BCG, its ellipticity and position angle (north of west). Second BCG properties are only listed if it was modelled separately from all other member galaxies.\\
{}$^a$ Measured manually in ds9. \\

}

\end{table*}

\begin{table*}
\caption{Thresholds used for running MIFAL}\label{mifalparams}
\centering
\begin{tabular}{c c c c c c c c}
\toprule \toprule
Cluster & Cluster &$\Delta r$ & $\Delta z$ &   $\Delta SB$ & min.  & max.  & $\chi^2_{ima}$\\
name & redshift & (\arcsec) &    & &images  &images  & cut\\
\midrule
M0329 & 0.450 & 5.2 & 0.5 & $300\%$ & 2 & 5 & 50\\
M1720 & 0.387 & 7.8 & 0.5 & $300\%$ & 2 & 5& 50\\
M1931 & 0.352 & 6.5 & 0.7 & $50\%$ & 2& 5& 50\\
\bottomrule
\end{tabular}                                   

{\footnotesize\flushleft
Note - Thresholds beyond which arcs are not considered potential multiple images for the results shown in \S \ref{SLanalysis}. See \S \ref{tempimagefile} for more details.\\
$Column~1~\&~2:$ cluster name and redshift; $Column~3:$ radius in arcseconds from a predicted counter image location within which candidate counter images are considered; $Column~4:$ photometric redshift threshold, within which multiple image candidates are considered; $Column~5:$ surface brightness threshold, in percent relative to the relensed arc, within which multiple image candidates are considered; $Column~6:$ minimum number of images allowed in a system; $Column~7:$ maximum number of images allowed in a system (most likely images are chosen); $Column~7:$ the $\chi^2_{ima}$  value below which an image is discarded from the final system list.\\
}

\end{table*}

\bsp
\label{lastpage}

\end{document}